%% file: mnras_template.tex
\documentclass[fleqn,usenatbib]{mnras}

\usepackage{newtxtext,newtxmath, booktabs}
\usepackage{amsmath}

\usepackage[T1]{fontenc}

\DeclareRobustCommand{\VAN}[3]{#2}
\let\VANthebibliography\thebibliography
\def\thebibliography{\DeclareRobustCommand{\VAN}[3]{##3}\VANthebibliography}

\usepackage{graphicx}	
\usepackage{amsmath}	

\title[The X-ray/UV correlation of QSOs]{Revisiting  the X-ray--to--UV relation of Quasars in the era of all-sky surveys}

\author[M. Chira et al.]{Maria Chira,$^{1}$\thanks{E-mail: mchira@noa.gr}
Antonis Georgakakis,$^{1}$
Angel Ruiz,$^{1}$
Shi-Jiang Chen,$^{2,3}$
Johannes Buchner,$^{2}$
Amy L. Rankine,$^{4}$
\newauthor
 Elias Kammoun,$^{5}$
Catarina Aydar,$^{2}$
Mara Salvato,$^{2}$
Andrea Merloni,$^{2}$
Mirko Krumpe$^{6}$
\\
$^{1}$Institute for Astronomy \& Astrophysics, National Observatory of Athens, V. Paulou \& I. Metaxa, 11532, Greece\\
$^{2}$Max Planck Institute for Extraterrestrial Physics, Giessenbachstrasse, 85741 Garching, Germany\\
$^{3}$School of Astronomy and Space Science, University of Science and Technology of China, Hefei 230026, People’s Republic of
China\\
$^{4}$ Institute for Astronomy, University of Edinburgh, Royal Observatory, Edinburgh EH9 3HJ, UK\\
$^{5}$ Cahill Center for Astrophysics, California Institute of Technology, 1216 East California Boulevard, Pasadena, CA 91125, USA\\
$^{6}$ Leibniz-Institut f\"ur Astrophysik Potsdam, An der Sternwarte 16, 14482 Potsdam, Germany\\
}

\date{Accepted XXX. Received YYY; in original form ZZZ}

\pubyear{\the\year{}}

\begin{document}
\label{firstpage}
\pagerange{\pageref{firstpage}--\pageref{lastpage}}
\maketitle

\begin{abstract}
The X-ray--to--UV relation of active galactic nuclei (AGNs), commonly parametrized via the monochromatic luminosities at 2500\,\AA\ and 2\,keV, reflects the energetic interplay between the accretion disc and the X-ray-emitting corona, and is key for understanding accretion physics. Previous studies suggest that disc-dominated emission becomes more prominent with increasing optical luminosity. However, the redshift evolution of this relation remains debated, and a dependence on Eddington ratio, predicted by accretion flow models, is still observationally unconstrained. We revisit this relation using a large, nearly all-sky sample by combining the SDSS DR16Q QSO catalogue with X-ray data from \textit{XMM-Newton} and the \textit{SRG/eROSITA} All-Sky Survey DR1, yielding 136{,}745 QSOs at redshifts $0.5 \leq z < 3.0$. We introduce a hierarchical Bayesian framework that treats X-ray detections and upper limits uniformly, enabling robust inference from both parametric and non-parametric models. We confirm a tight, sublinear $\log L_X({\rm 2\,keV})$--$\log L_{\nu}({\rm 2500\,\mathring{A}})$ correlation, but with a normalization at the lower end of previous estimates. Contrary to most literature results, we detect a mild but systematic redshift evolution: the relation flattens and its intrinsic scatter decreases at higher redshift. This trend is consistent with disc emission increasingly dominated by scattering and enhanced energy transfer to the X-ray corona, potentially indicating redshift evolution in the X-ray bolometric correction. We find no significant dependence on Eddington ratio, in tension with recent accretion flow models.
\end{abstract}



\input{Introduction}
\input{Data_v2}
\input{Method_restructured}
\input{Analysis_restructured}

\input{Discussion_v1}
\input{Conclusions}

 \section*{Acknowledgements}
The authors wish to thank the anonymous referee for their careful reading of the paper and their insightful comments. The research leading to these results has received funding from the Hellenic Foundation for Research and Innovation (HFRI) project "4MOVE-U" grant agreement 2688, which is part of the programme "2nd Call for HFRI Research Projects to support Faculty Members and Researchers". This work is based on data from eROSITA, the soft X-ray instrument aboard SRG, a joint Russian-German science mission supported by the Russian Space Agency (Roskosmos), in the interests of the Russian Academy of Sciences represented by its Space Research Institute (IKI), and the Deutsches Zentrum für Luft- und Raumfahrt (DLR). The SRG spacecraft was built by Lavochkin Association (NPOL) and its subcontractors, and is operated by NPOL with support from the Max Planck Institute for Extraterrestrial Physics (MPE). The development and construction of the eROSITA X-ray instrument was led by MPE, with contributions from the Dr. Karl Remeis Observatory Bamberg \& ECAP (FAU Erlangen-Nuernberg), the University of Hamburg Observatory, the Leibniz Institute for Astrophysics Potsdam (AIP), and the Institute for Astronomy and Astrophysics of the University of Tübingen, with the support of DLR and the Max Planck Society. The Argelander Institute for Astronomy of the University of Bonn and the Ludwig Maximilians Universität Munich also participated in the science preparation for eROSITA. The Argelander Institute for Astronomy of the University of Bonn and the Ludwig Maximilians Universität Munich also participated in the science preparation for eROSITA. The eROSITA data shown here were processed using the eSASS/NRTA software system developed by the German eROSITA consortium. ALR acknowledges support from a Leverhulme Trust Early Career Fellowship. For the purpose of open access, the author
has applied a Creative Commons Attribution (CC BY) licence to any Author
Accepted Manuscript version arising from this submission. MK is supported by DLR grant FKZ 50 OR 2307. This research made use of  Astropy,\footnote{\href{www.astropy.org}{www.astropy.org}} a community-developed core Python package for Astronomy \citep{astropy:2013, astropy:2018}. 

\section*{Data Availability}

The data and code used in this paper are publicly available at \href{doi.org/10.5281/zenodo.17476605}{doi.org/10.5281/zenodo.17476605}.

%
%
\bibliographystyle{mnras}
\bibliography{LXLUV}

\input{Appendix_A}

\bsp	
\label{lastpage}
\end{document}

%% file: Introduction.tex
\section{Introduction}\label{sec:intro}

Active Galactic Nuclei (AGN) are among the most luminous and energetic objects in the universe, powered by accretion onto supermassive black holes \citep[SMBHs;][]{LyndenBell1969, Rees1984}. Their spectral energy distributions (SEDs) span a wide range of wavelengths, providing crucial insights into the physical processes governing their accretion dynamics. The main components of an accretion flow include an accretion disc that emits predominantly in the ultraviolet \citep[UV,][]{Shakura1973,Koratkar1999, Davis2007}, a hot corona responsible for X-ray emission via inverse Compton scattering \citep{haardt1991, Haardt1993}, and perhaps a warm corona that may bridge the energy gap between the two \citep{Petrucci2018} and account for the frequently encountered soft excess \citep[e.g.,][]{Chen2025}.

The currently available observational data reveal differences in the SEDs and properties of AGN, pointing to the need for sophisticated and versatile models of accretion. The most basic toy-model which explains reasonably well the dominant parts of the AGN emission is that the gravitational energy of the infalling material onto the SMBH is transformed into radiation that peaks at the UV part of the electromagnetic spectrum. It is these seed UV photons from the disc that are reprocessed via inverse Compton scattering in the hot corona and re-emitted in the X-rays. This simplistic yet solid schema for the energetic interplay between the disc and the corona should be reflected on the SEDs of AGN populations, resulting into a correlation of the X-ray and UV radiation of AGN.   

Indeed, such a correlation was revealed already since the first X-ray surveys became available and is currently well established as a fundamental universal property of AGN. The correlation is found to be well described by a linear relation between the monochromatic luminosities $\log{L_X(\rm 2\,keV)}$ and $\log{L_\nu(2500\mathring{\rm A})}$ with a slope that is flatter than unity and is typically found to be $\approx0.6$ \citep[e.g.,][]{Tananbaum1979, Tananbaum1986, Strateva2005,Steffen2006, LR2015, Arcodia2019, LR2023}. Such a correlation implies that the more UV luminous an AGN is, the more disc dominated its SED is. An alternative equivalent parametrization of the X-ray--to--UV correlation commonly utilized in the literature involves the spectral index $\alpha_{OX}$ \citep{Tananbaum1979}, which quantifies the relative intensity of the monochromatic X-ray luminosity at $2\,{\rm keV}$ with respect to the monochromatic $2500\rm{\mathring{A}}$ UV emission \citep[e.g.,][]{JUST2007, Lusso2010}. The observed X-ray-to-UV correlation and the underlying mechanism responsible for it is a key element towards the deeper understanding of the accretion process in AGN and can serve as an informative constrain of accretion models \citep{Arcodia2019, LR2017}. 


The X-ray--to--UV correlation also shows scatter, which is typically of the order of $0.4 \text{ dex}$. This may point to diverse accretion flow physical
conditions and/or stages but may also be related to observational
effects and unaccounted host contribution. Several studies explored potential contributors to the scatter including differences in black holes masses or Eddington ratios, AGN variability, absorption effects, diverse inclination angles of the accretion disc relative to the observer, biases in X-ray flux estimates or non-simultaneity of X-ray and UV observations \citep[e.g.,][]{Kelly2007, LR2017, Chiaraluce2018, Bisogni2021, Signorini2024}. However, a fraction of the scatter could simply reflect a level of intrinsic diversity among AGN populations which may contribute to deviations from the mean expected trend. From this point of view, the relation can serve as a tool for the identification of possibly interesting "extremities" in terms of X-ray luminosity with respect to the average population \citep[e.g.,][]{Laurenti2024, Pu2020}. The construction of a "clean" quasar (QSO) sample for cosmological purposes leads to a lower dispersion of the order of $\sim 0.24\text{\,dex}$ \citep[e.g.,][]{LR2015, LR2016, LR2020}. Moreover, \citet{Signorini2024}, based on mock samples, deduced that the intrinsic scatter of the relation for such a "clean" sample is of the order of $\sim 0.06 \text{\,dex}$, with the main contributors to the additional dispersion being the AGN variability, while the use of X-ray photometric rather than X-ray spectral data has minimal contribution to the scatter \citep{Signorini2024}. 
  
An important open question is whether the X-ray--to--UV correlation evolves with redshift. A common finding in the literature is that the scatter decreases with redshift \citep[e.g.][]{Li2021, Rankine2024}. During the previous decades, the majority of authors report no evidence for a redshift dependence of the correlation in terms of slope or normalization \citep[]{JUST2007, LR2015,LR2016}. However, there have been exceptions \citep[e.g.,][]{Shen2006, Kelly2007} and a number of recent studies \citep{Rankine2024,Wang2022_copulas, Khadka2022} report evidence for redshift evolution, potentially suggesting changes in the structure or efficiency of AGN accretion with cosmic time. The implications of such an evolution, if confirmed, could have profound consequences for AGN feedback and galaxy/SMBH coevolution scenarios, cosmic structure formation, and even the use of AGN as cosmological probes \citep{LR2019}, under the assumption that the X-ray--to--UV is a global property of AGN and does not evolve with redshift.

Observational biases, such as flux-limited samples, obscuration, and instrumental sensitivities, can distort the inferred relationship and contribute to discrepancies between different studies. Addressing these biases is crucial for obtaining a robust understanding of AGN accretion physics and ensuring accurate interpretations of their emission properties across cosmic time. Another important point towards an unbiased sample is the proper inclusion of upper limits which poses significant challenges in studying the X-ray to UV correlation. This is essential to incorporate the total information currently available from recent observations and especially the SRG/eROSITA All-Sky X-ray survey \citep{Merloni2024, Predhel2021}, which opens a new window for studying the global properties of AGN. In previous studies, different authors have followed different approaches regarding the X-ray non-detected QSOs, which could be a contributing factor in the deviations between their results. Specifically, some works select their samples based exclusively on X-ray detections \citep[e.g.,][]{LR2017}, others \citep[e.g.,][]{Steffen2006,Green2009} tailor their selection criteria to ensure small contribution of upper limits in their samples, while others \citep[e.g.,][]{Rankine2024,Timlin2021,Vignali2003} include upper limits in their samples, inducing a methodological challenge in their analysis.

In this work, we present a new Bayesian methodology to ensure a uniform treatment of X-ray detections and upper limits. We combine Sloan Digital Sky Survey Quasar Catalog data release 16 \cite[SDSS DR16Q][]{Lyke2020} with the recently available German eROSITA All-Sky Survey Data Release 1 \citep[eRASS-DE DR1;][]{Merloni2024} and the {\it XMM-Newton} archive. DR16Q includes several epochs of optical observations which correspond to different targeting strategies and magnitude limits, which softens the strong luminosity vs redshift bias of a sample with a single flux limit. The combination with the latest eROSITA All-Sky Survey X-ray data, as well as the more sensitive {\it XMM-Newton} observations (median exposure time of 10\,ks vs 0.1-0.2\,ks for eRASS1), results in a sample of unprecedented size and maximizes the dynamical range of the QSO properties, enabling a detailed and robust study of the X-ray--to--UV correlation and its dependencies.

In Section~\ref{sec_Data}, we present the different data samples combined and the selection procedure to construct the QSO sample analysed in this work, as well as the approach for calculating the monochromatic luminosity at $2500\rm \mathring{A}$. Section~\ref{sec_Method} presents our Hierarchical Bayesian methodology and discusses the different model flavors adopted throughout this analysis. Section~\ref{results} presents our results on the correlation and its dependence on redshift and other QSO properties, such as the Eddington ratio, and performs a comparison with the predictions of the model of \citet{Kubota_Done2018}. In Section~\ref{sec:discussion} we discuss the interpretation of our main results, their implications on our current understanding of the accretion processes, also within the framework of the {\sc kynsed} model of \citet{Dovciak2022}, and discuss the properties of a subpopulation of X-ray-faint sources as identified based on their deviation from our inferred relation. Finally, we summarize our conclusions in Section~\ref{conclusions}.  Throughout this paper we assume a standard $\Lambda$CDM cosmology ($\Omega_m = 0.3, H_0 = 70 \rm km/s/Mpc$) .

%% file: Data_v2.tex
\section{Data}\label{sec_Data}

\subsection{Optical Quasar Catalogue: DR16Q}
The QSO
sample used in this work is a subset of the Sloan Digital Sky Survey Quasar Catalog data release 16 \cite[SDSS DR16Q;][]{Lyke2020}. It contains 750,414 spectroscopically confirmed quasars with spectroscopic and multiwavelength photometric information and represents the final quasar catalog from SDSS-IV. It brings together quasar spectra from several stages of the Sloan Digital Sky Survey (SDSS), creating one of the most complete samples of quasars available across a wide range of redshifts. It includes sources observed during SDSS-I and SDSS-II, which mainly focused on bright quasars at low redshift $(z \lesssim 2.2)$ as part of the Legacy survey \citep{York2000, Schneider2010}. The catalog also contains data from SDSS-III BOSS survey \citep{Dawson2013}, which added many high-redshift quasars for studies of the Lyman-$\alpha$ forest and cosmology. The largest new contribution comes from SDSS-IV eBOSS program \citep{Dawson2016}, which filled in the redshift gap between $z \approx 0.8 - 2.2$, and also continued to observe quasars at higher redshifts $(z \gtrsim 2.1)$. In addition, DR16Q includes quasars from smaller programs such as SPIDERS \citep[X-ray-selected AGN, ][]{Dwelly2017, Comparat2020} and TDSS \citep[quasars selected by their variability, ][]{Morganson2015, MacLeod2018}, as long as their spectra confirmed a quasar classification. The physical properties for these QSOs, such as black hole masses, Eddington ratios and bolometric luminosities, are from \cite{Wu_Shen2022}. 

The DR16Q catalogue contains 1{,}011 QSOs for which the PSFFLUX/PSFMAG parameters are flagged with a value of $-9999$ in all five SDSS bands. For the vast majority of these sources this is because they are selected as targets from external catalogues and therefore Sloan photometry is not available for them \citep[e.g. see Section 8 of][]{Lyke2020}. A minority are located close to very bright stars that clearly contaminate their photometry. These 1{,}011 sources are removed from our analysis. As discussed previously, the DR16Q includes QSOs targeted using different selection criteria. In this paper, we limit the sample to QSOs selected based on their optical and/or mid-infrared colours, i.e. that show blue rest-frame UV/optical continua. These correspond to the \texttt{CORE} QSO targeting  criteria of the SDSS-I to V. In practice, we select QSOs that belong to target classes \texttt{QSO\_CORE\_MAIN} of SDSS-III/BOSS \citep{Dawson2013}, \texttt{QSO1\_EBOSS\_CORE}, \texttt{QSO\_EBOSS\_CORE} of SDSS-IV/EBOSS \citep{Dawson2016}, as well as any sources that fulfill the above selection criteria but had been targeted by the earlier SDSS-I and II programmes. This filtering is performed based on  the targeting bitmasks of SDSS catalogues, i.e. bit 17 of \texttt{EBOSS\_TARGET1} and \texttt{EBOSS\_TARGET0} (select SDSS I/II QSOs), bit 40 of mask \texttt{BOSS\_TARGET1} (selects \texttt{QSO\_CORE\_MAIN}), bit 10  of \texttt{EBOSS\_TARGET0} (selects \texttt{QSO\_EBOSS\_CORE}), and bit 40 of \texttt{EBOSS\_TARGET1} (\texttt{QSO1\_EBOSS\_CORE}). This selection contains 534{,}322 QSOs.

 Moreover, we limit our sample to the redshift range $0.5\leq z<3.0 $ corresponding to 503{,}182 QSOs. The lower limit is imposed to avoid possible heavy host galaxy contamination, while the upper limit ensures trustworthy $L_\nu (\rm 2500\mathring{A})$ calculations. 

\subsection{X-ray Data}

A key objective of our analysis is to constrain the $L_X({\rm 2\,keV}) - L_\nu({\rm 2500\mathring{A}})$ correlation of SDSS QSOs and explore its possible evolution with redshift. The X-ray data for the DR16Q sample are from the {\it XMM-Newton} archive and the German eROSITA All Sky Survey Data Release 1 \citep[eRASS-DE DR1;][]{Merloni2024}. The latter dataset consists of observations carried out in the first six months of the SRG/eROSITA all-sky survey (eRASS1) whose proprietary rights lie with the German eROSITA consortium (eROSITA-DE). As explained in Section~\ref{sec_Method}, the Bayesian statistical methodology developed to model the $L_X({\rm 2\,keV}) - L_\nu({\rm 2500\mathring{A}})$ correlation does not rely only on QSOs individually detected on the {\it XMM-Newton} archival observations or the eRASS-DE DR1. Instead, our modelling approach uniformly treats X-ray detections and upper limits by using X-ray aperture photometry information extracted at the positions of DR16Q QSOs with available X-ray data. 

\subsubsection{\bf XMM-Newton} For QSOs in the SDSS DR16Q catalogue within the XMM-Newton footprint, we obtain photometry products in the $0.2-2\, \rm keV$ energy band (X-ray counts, vignetting-corrected exposure times, background levels) by querying the RapidXMM Upper Limit Server \citep{Ruiz2022} at the optical position of each QSO. Our analysis uses the EPIC-PN detector \citep{Struder2001} only,  because of its higher sensitivity. The RapidXMM aperture size is fixed to 15\,arcsec, corresponding roughly to 70\% of the Encircled Energy Fraction (EEF) of the {\it XMM-Newton} PSF \citep[see][for more details]{Ruiz2022}. The query initially returns 27,081 {\it XMM-Newton} photometric measurements. We remove 3,531 of them because they are affected by high-particle background levels. We further remove 2,527 photometric measurements because the corresponding QSO position lies close to the edges of the {\it XMM-Newton} field of view or CCD gaps of the EPIC \citep[European Photon Imaging Camera;][]{Turner2001} detectors, according to the flagging system of RapidXMM \citep[see Table 2 of][]{Ruiz2022}. Moreover, QSOs with potential photon contamination from the PSF wings of nearby sources need to be excluded from the analysis. For this purpose, the optical positions of QSOs are cross-matched with the data release 14 of the 4XMM  serendipitous X-ray source catalogue \citep{Webb2020} within a radius of $6\,\text{arcsec}$. This yields a total of 11,049 associations. For the density of the 4XMM-DR14 X-ray detections (692,109 unique sources over an area of 1,383 $\rm{deg^2}$) and the adopted matching radius, the expected chance association rate is 0.44 per cent. We next find 820 QSOs in the DR16Q catalogue that lie with an annulus with inner and outer radii of $6\, \text{arcsec}$ and $45\, \text{arcsec}$ respectively from 4XMM-DR14 X-ray sources. We assume that these QSOs are not associated with the nearby X-ray source. The X-ray photometry of these sources may be contaminated by the photons in the PSF wings of the nearby 4XMM-DR14 detection and are therefore removed from our sample. This filtering results in 21,199 X-ray photometric measurements. In the case of multiple {\it XMM-Newton} observations of a given QSO we sum their X-ray photon counts, expected background levels and exposures of the individual observations. This results to 13,933 unique SDSS QSOs with X-ray aperture photometric information from the RapidXMM database.

 \subsubsection{\it \bf  eROSITA} In the case of eRASS-DE DR1 we query the respective X-ray photometric information in the $0.2-2.3 \,\rm keV$ energy band at the optical positions of the selected DR16Q QSOs  using the {\sc apetool} task of the eROSITA Science Analysis Software System \cite[eSASS;][]{Brunner2022}. Photons are extracted within an aperture that correspond to the  75\% EEF of the eROSITA PSF at each optical position. This returns photometric measurements for 134,309 unique QSOs. Similar to the {\it XMM-Newton} approach we remove from the analysis QSOs whose aperture photometry may be contaminated by the photons in the PSF wings on nearby X-ray detections. First we identify all QSOs within a distance of 45\,arcsec from eRASS1 X-ray sources with detection likelihood parameter in the eRASS1 catalogue {\sc det\_ml}$> 7$ \citep[see][for the expected false positive fraction as a function  {\sc det\_ml}]{Seppi2022}. Among these QSOs there are true associations with the eRASS1 sources. We use the eRASS1 optical identification catalogue \citep{Salvato2025}  to identify a total of 17,249 unique DR16 SDSS QSOs that are securely associated with an eRASS1 X-ray detection, while 2033 sources are found within 45\,arcsec from an eRASS1 X-ray source but are not associated with it. This step removes 1,314 QSOs from our sample. These sources may have their X-ray photometry contaminated by the PSF wings of nearby X-ray detections.
 
 \subsection*{}There is an overlap of 4,236 QSOs between the {\it XMM-Newton} and eROSITA samples. For these sources, we use the deeper {\it XMM-Newton} observations. Combining the two samples leads to a total number of 142{,}692 QSOs ({\it XMM-Newton}: 13,933, eROSITA: 128,759). 

\subsection{Additional Filtering}

Further selection criteria are applied to avoid contamination from X-ray radiation emitted from the gas of galaxy clusters. We exclude from our sample QSOs found in the vicinity of known X-ray clusters from the ROSAT All-sky Survey \citep[RXGCC,][]{Xu2022}, the XMM Cluster Survey \citep[XCS,][]{Mehrtens2012}, the XMM CLuster Archive Super Survey \citep[X-CLASS,][]{Clerc2012}, the XMM-XXL survey \citep[XXL365,][]{Adami2018}, and the eROSITA Final depth Equatorial Survey \citep[eFEDS,][]{Liu2022}. The exclusion radius of each individual cluster is empirically determined and is different for the {\it XMM-Newton} and eROSITA observations, due to the different PSF of the telescopes. For details on the exclusion radii calculations we refer the reader to \citet[][their Section 2]{Georgakakis2024}. This step excludes 2,029 QSOs. 

 The significance of the X-ray--to--UV luminosity ratio of QSOs is that it provides a diagnostic of the interplay between the accretion disc and the hot X-ray emitting corona. In the case of Blazars and more generally radio loud QSOs, jets may contribute significantly to the flux at X-ray energies. The X-ray luminosity of these systems therefore provides limited information on the hot corona. We remove such QSOs from the analysis by estimating the radio loudness parameter \citep{Kellermann1989} 

 \begin{equation}
   R_{F} = \frac{F_\nu (\rm 5\,GHz)}{F_\nu(\rm 4500\mathring{A})},
 \end{equation}\label{eq:Rf}
 
\noindent where $F_\nu(\rm 5\,GHz)$ and $F_\nu(\rm 4500\mathring{A})$ are the flux densities at 5\,GHz and  $4500\mathring{A}$, respectively. To estimate the flux density at $5\,\mathrm{GHz}$, we utilize three large-area radio surveys: the FIRST \citep[Faint Images of the Radio Sky at Twenty Centimeters;][]{Becker1995} survey, the second data release of LoTSS \citep[LOFAR Two-meter Sky Survey;][]{Shimwell2017, Shimwell2022}, and the CRATES-FSRS \citep[Combined Radio All-Sky Targeted Eight-GHz Survey - Flat-Spectrum Radio Source Catalog;][]{Healey2007}, which is compiled from existing observations.

The FIRST survey covers approximately $10{,}000\, \mathrm{deg}^2$, including most of the SDSS area, with a median $1\sigma$ RMS sensitivity of $0.141 \, \mathrm{mJy \, beam}^{-1}$ at $1.4 \, \mathrm{GHz}$. The LoTSS DR2 spans about 5,700 $\mathrm{deg}^2$ at 144 $\mathrm{MHz}$ and achieves a deeper flux density limit compared to FIRST, with a median $1\sigma$ RMS sensitivity of 83 $\mu \mathrm{Jy \, beam}^{-1}$. The SDSS DR16Q quasar catalogue has already been cross-matched with the FIRST source positions. For QSOs lacking FIRST counterparts, we compute $3\sigma$ upper limits to their $1.4 \, \mathrm{GHz}$ flux densities using the FIRST RMS noise maps and applying the relation $0.25 + 3\sigma_\mathrm{RMS}$, where $\sigma_\mathrm{RMS}$ is the FIRST RMS noise at the source position and $0.25$ accounts for the CLEAN bias correction \citep{White1997}.

The association of SDSS DR16Q QSOs with LoTSS sources is based on the optical counterpart positions provided by \citet{Hardcastle2023}. For QSOs without LoTSS matches, we adopt a $3\sigma$ upper limit to their $144 \, \mathrm{MHz}$ flux densities of $0.3 \, \mathrm{mJy}$, derived from the median RMS noise of $83 \, \mu \mathrm{Jy \, beam}^{-1}$. In both radio surveys, flux densities are converted to rest-frame values at $5 \, \mathrm{GHz}$ assuming a power-law spectral index of $-0.8$  \citep[$F_\nu \propto \nu^{-0.8}$;][]{Sanchez2018}.

Additionally, SDSS QSOs are cross-matched with the CRATES-FSRS catalogue, which includes over 11,000 bright, flat-spectrum radio sources at $8.4 \, \mathrm{GHz}$. The CRATES-FSRS catalogue is augmented with reprocessed archival VLA and ATCA data, as well as additional observations to address coverage gaps. QSOs with a CRATES counterpart are excluded from the analysis.

A common choice for defining radio loud sources based on the $R_{F}$ for AGN is
$R_F < 10$ \citep{Kellermann1989}. However, current observational evidence suggests that the X-ray flux of AGN is dominated by the corona up to $R_F \sim 100$ \citep[e.g.,][]{Zhu2020}. In this work, we select a moderate threshold of $R_F = 30$. This step filters out $3{,}911$ QSOs. 

Finally, we remove 7 QSOs with extremely high X-ray photon counts ($>10^4$) to avoid pile-up effects. The final sample contains 136,745 QSOs of which 13,412 have XMM-Newton and 123,333 have eROSITA data associated with them.

\subsection{SED fitting: calculation of the UV luminosity.}

The monochromatic luminosity at 2500\AA, $L_\nu (\rm 2500\mathring{A})$, is derived by fitting template models to the observed multi-band Spectral Energy Distribution (SED) of SDSS QSOs, incorporating near-infrared, optical, and ultraviolet photometric data. The SED fitting procedure is selected since it yields the intrinsic $L_\nu(\rm 2500 \mathring{A})$, corrected for host galaxy extinction. Since the optical/UV radiation is more strongly affected by extinction than the more penetrating X-rays, using the intrinsic optical/UV luminosity is preferable for our purposes. The DR16Q catalog includes multiwavelength data, either by direct cross-matching or via force-photometry. Among others, sources cross-matched to data from the Galaxy Evolution Explorer \citep[GALEX,][]{Martin2005} and the UKIRT Infrared Deep Sky Survey \citep[UKIDSS][]{Lawrence2007}. Specifically, we use the GALEX/NUV band ($350-1750 \mathring{A}$), which includes positive fluxes for 552,025 SDSS QSOs, along with four UKIDSS bands: Y ($1.02 \mu m$), J ($1.25 \mu m$), H ($1.63 \mu m$), and K ($2.20 \mu m$). The corresponding number of sources with positive fluxes for each UKIDSS band are 150,147, 149,629, 149,502, and 150,288, respectively, with 146,500 sources showing positive fluxes in all four bands. We avoid the inclusion of FUV data for this exercise, due to the low detection rate and
complex IGM absorption \citep[e.g.][]{Cai2023}. Additionally, we cross-matched the DR16Q catalog with the VISTA Hemisphere Survey \citep[VHS;][]{VHS2013} DR6, using a cross-match radius of $0.5$ arcseconds. This resulted in 35,431 SDSS QSOs with positive flux in at least one of the four VHS bands (Y: $1.02 \mu m$, J: $1.25 \mu m$, H: $1.64 \mu m$, Ks: $2.15 \mu m$). For the
density of the VHS-DR6 detections (1,374,207,485 sources over an area of 14,486 $\rm{deg^2}$) and the adopted matching
radius the expected chance-association rate is 0.58 per cent.

SED fitting was performed using the Code Investigating GALaxy Emission \citep[CIGALE;][]{Boquien2019}, a widely used multi-component SED fitting algorithm. Following the recommendations of \citet{Yang2020} for the SED fitting of QSOs, we utilized two modules: `skirtor2016' for the AGN component, and `redshifting' to account for the redshift of the SED. The `skirtor2016' module is based on the AGN models of \citet{Stalevski2016} and employs a realistic two-phase clumpy torus model\footnote{The two-phase clumpy torus model describes the AGN torus as a mixture of dense dusty clumps and a diffuse interclump medium, providing a more realistic framework for explaining its infrared emission and anisotropic obscuration.}. A list of the CIGALE modules and the grid of parameters used for this analysis is provided in Table \ref{t1}. Although the assumption that the AGN component dominates the SED of SDSS QSOs is reasonable, the stellar emission from the host galaxy may become non-negligible at lower luminosities and/or redshifts. In principle, this can be addressed by decomposing the SED into stellar and non-thermal nuclear emission from the AGN by e.g. adding  galaxy templates into CIGALE. In practice however, such a decomposition exercise is non-trivial particularly in the case of broad-band photometry. Covariances between model parameters and systematic effects may lead to significant uncertainties in the decomposition results. The optical spectra instead, contain a number of features (e.g. continuum shape, spectral breaks, absorption lines) that allow a better handle on the host galaxy contribution to the observed emission \citep[e.g.][]{Wu_Shen2022, Ren2024}. We therefore choose to use the host-galaxy fractions determined by \cite{Ren2024} for SDSS DR14Q QSOs at $z<0.8$ to correct the $L_\nu(\rm 2500 \mathring{A})$ estimated from the SED fitting. At $z\approx 0.8$  and beyond the host galaxy contribution to the DRQ14 QSO continuum becomes negligible. In practice, we estimate the host galaxy fraction at 2500\AA\, using the \cite{Ren2024} results and then reduce the inferred $L_\nu(\rm 2500 \mathring{A})$ accordingly. The median decrease factor applied to $L_\nu(\rm 2500 \mathring{A})$ is 1.6.

\begin{figure}
   \centering
   \includegraphics[width=1.0\columnwidth]{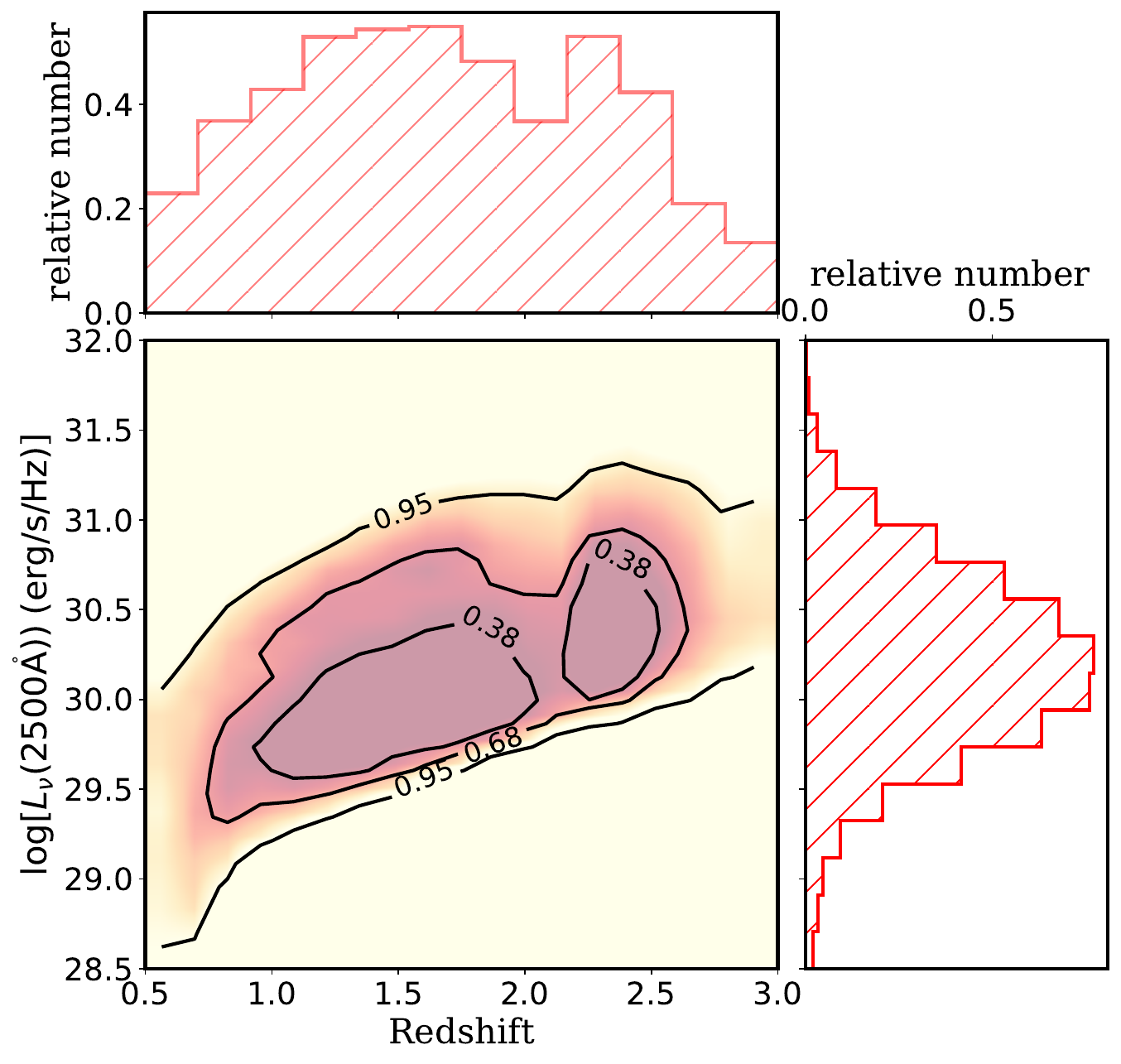}
      \caption{Distribution of the DR16Q QSO sample on the $\log{L_\nu(\mathrm{2500\mathring{A}})} - z$ plane. Projected frequency histograms of the redshift and the $\rm 2500\mathring{A}$ luminosity are presented in the barred plots on the horizontal and vertical axis respectively. 
}
         \label{LUV_Z}
   \end{figure}

The SED fitting procedure provides an estimate of the intrinsic luminosity, $L_\nu(\rm 2500 \mathring{A})$, as well as the extinction in the polar direction, $E(B-V)$. The results of the SED fitting are shown in Figure \ref{LUV_Z}, which plots the distribution of the QSO sample on the $z- \log{L_\nu{2500\mathring{A}}}$ plane. To validate our methodology, we also estimate $L_\nu(\rm 2500 \mathring{A})$ 
by applying linear interpolation/extrapolation based on the available photometric fluxes at the redshifted wavelength $2500{\rm \mathring{A}} \cdot (1+z)$. A detailed description of these methods, along with a comparison showing good agreement between the results, is provided in Appendix A.


%% file: Method_restructured.tex
\section{Hierarchical Bayesian Methodology}\label{sec_Method}

A sizeable fraction of the QSOs in our sample are not formally detected in either eROSITA ($\sim {95\%}$ upper limits) or {\it XMM-Newton} ($\sim 32\%$ upper limits) and, as a result, only X-ray upper limits are available for them. Nevertheless, these non-detections often provide useful constraints on the X-ray/UV properties of QSOs and it is therefore important to take them into account. Moreover, ignoring upper limits may bias the inferred relation between UV and X-ray luminosities. It is thus essential to develop a robust statistical methodology that (i) encapsulates the Poisson nature of the X-ray photometric data, (ii) allows the homogeneous treatment of both X-ray detections and upper limits, and (iii) captures the covariances between model parameters. We choose to develop a new hierarchical Bayesian framework that fulfills the requirements above and at the same time facilitates the use of complex likelihood functions and the integration of prior information into the analysis. These features are crucial for capturing the nuanced correlations between X-ray and UV luminosities of QSOs, based on multiparametric data.

Our approach relies on X-ray aperture photometry to address the point above. This is because the extracted photon counts in this case follow the Poisson distribution, which facilitates the inference calculations. Suppose the $i$-th QSO in the sample with redshift $z$ and a monochormatic luminosity $L_X(\rm 2keV)$. For this individual source, the probability of observing a number of X-ray counts, $N_i$, is described as 

\begin{align}\label{eq:poisson-basic}
P(N_i| \lambda _i) = \frac{{\rm e}^{- \lambda _i} \lambda _i^{N_i}}{N_i!},
\end{align}

\noindent where $\lambda_i$ is the expectation value of the Poisson that can be expressed as

\begin{align}\label{eq:poisson-expectation}
    \lambda_i = f_{X,\;i}  \cdot EEF_{i} \cdot t_i \cdot ECF_i + B_i,
\end{align}

 \noindent where $f_{X,\;i}$ is the source X-ray flux in a given energy band, $B_i$ is the background level within the aperture, $ECF_i$ is the energy conversion factor that depends on the spectral model of the source and the characteristics of the detector, $t_i$ is the vignetting-corrected exposure time, and $EEF_i$ is the Enclosed Energy Fraction within the aperture. The conversion from monochromatic luminosity $L_X(\rm 2keV)$ to flux depends on the redshift (i.e., cosmological luminosity distance) and the source's X-ray spectral shape. For the latter, we assume a power-law spectrum with index $\Gamma=1.9$\footnote{Assuming a steeper spectral slope of $\Gamma=2.3$, typical for QSOs with strong soft excess component \citep[e.g.][]{Chen2025}, increases the inferred $L_X(\rm 2keV)$ by about 5\%.} that is absorbed by the Galactic hydrogen column density, $N_H$, in the direction of each QSO, based on the maps of \cite{Kalberla}, using the python module {\sc gdpyc} \citep{angel_ruiz_2018_1482888} \footnote{\href{https://github.com/ruizca/gdpyc}{https://github.com/ruizca/gdpyc}}. Therefore, for a single object, studied independently of the population, the likelihood of the source having monochromatic luminosity $L_X(\rm 2keV)$ given the observations can be written as 
 
\begin{align}\label{eq:single_likelihood}
  \mathcal{L}_i= P(N_i \; | \; \lambda_i)\cdot P\left[ \lambda_i\;|L_X(\rm 2keV) \right],
\end{align}

\noindent where $P\left[\lambda_i|L_X(\rm 2keV)\right]$ is the probability of the 
expectation value given $L_X(\rm 2keV)$. At the population level we know that the X-
ray and UV luminosities of QSOs are correlated and we can therefore use this fact to inform the likelihood of individual sources above. We therefore introduce another layer in the likelihood that links the X-ray and UV luminosities of QSOs assuming some model distribution

\begin{align}\label{eq:model-gh}
L_X(\rm 2keV) \sim \mathcal{M}(\Theta),
\end{align}

\noindent where the tilde signifies that the quantity on the left-hand side of the equation follows a distribution dictated by the model of choice on the right-hand side, $\mathcal{M}(\Theta)$. The set of parameters $\Theta$ generically describe the various dependence of $L_X(\rm 2keV)$ on, e.g., redshift or UV luminosity. It also encapsulates any possible functional form of the above parameters, e.g., linear correlation. Based on the above, the likelihood of the single object is modified as 

\begin{align}\label{eq:multi_likelihood}
  \mathcal{L}_i= P(N_i|\lambda_i)\; P(\lambda_i|\mathcal{M}),
\end{align}

\noindent where $P(N_i|\lambda_i)$ is the individual source probability for $N_i$ counts given an expectation value $\lambda_i$ and $P(\lambda_i|\mathcal{M})$ is the probability of that expectation value given the population-level properties described by model $\mathcal{M}$. Finally, the total likelihood of all sources in a sample is the product of the single-object likelihoods

\begin{align}\label{eq:likelihood}
    \mathcal{L} = \prod_{i} \mathcal{L}_i.
\end{align}

\noindent Next we consider four different options for the model distribution $\mathcal{M}(\Theta)$. First, we explore the X-ray--to--UV correlation in a non-parametric manner by not enforcing an explicit correlation between $L_X(\rm 2keV)$ and $L_\nu(\rm 2500 \mathring{A})$. For this purpose, we bin the sample into relatively narrow bins of redshift, $\Delta z=0.5$, and UV luminosity, $\Delta\log L_\nu(\rm 2500 \mathring{A})=0.5$, and then infer the $\log L_X(\rm 2keV)$ distribution of the sample within these bins assuming it is described by a Gaussian. Put differently, the adopted model, $\mathcal{M}(\Theta)$, in this case is formulated as   

\begin{align}\label{equation:model-flavour-1}
   \mathcal{M}(\Theta) =  \mathcal{N}(\mu,\;\sigma).
\end{align}

\noindent Where the symbol $\mathcal{N}$ denotes the normal distribution with a mean $\mu$ and a scatter $\sigma$.  Both these parameters are a function of both redshift and UV luminosity. Although they are assumed to be constant at fixed redshift and UV-luminosity bin, they are allowed to vary across bins. In this particular flavour of the model the number of free parameters in the fit are $2\times N_{bins}$, where $N_{bins}$ is the adopted number of redshift and UV-luminosity bins. We note that throughout our analysis and all model versions $\mu = \log{L_X(\rm2keV)}$.

The next model flavour is tailored to allow us to explore in a least parametric manner possible redshift evolution effects of the $\log L_X (\mathrm{2keV})$ distribution within narrow intervals of $\log L_\nu (\mathrm{2500\, \mathring{A}})$. We assume the normal distribution of Equation \ref{equation:model-flavour-1} but in this case the mean parameter is given by the relation

\begin{align}\label{equation:model-flavour-2}
   \mu(z) = \mu(z_0) + \Delta\mu(z).
\end{align}

\noindent Therefore, we explore at fixed $\log L_\nu (\mathrm{2500 \, \mathring{A}})$ possible variations of the parameter $\mu(z)$ at redshift $z$ relative to $\mu(z_0)$ at a reference redshift of $z_0$. Put differently, the  $\mu(z_0)$ is a global parameter shared across all redshift bins within a given $\log L_\nu (\mathrm{2500\, \mathring{A}})$ interval. This allows us to capture evolutionary effects by simultaneously fitting the model to all QSOs at fixed UV luminosity bin. In this modeling approach, the scatter parameter, $\sigma$, of Equation \ref{equation:model-flavour-1} is assumed to be independent of redshift (although it is different for each UV luminosity bin). In practice, we use the same redshift and UV luminosity intervals as in the first model flavour and set the reference redshift bin to the interval $z=1.0-1.5$. This bin is chosen arbitrarily as reference because it is among the most populated ones in our sample. The free parameters in this case are the $\mu(z_0)$, $\sigma$ for each UV luminosity interval and the corresponding $\Delta\mu(z)$ for the each redshift and  luminosity bin. The total number is $(N_{z-bins} +1)\cdot N_{UV-bins}$, where $N_{z-bins}$, $N_{UV-bins}$ are the number of redshift and UV luminosity bins, respectively.

The next model version follows conventions in the literature and adopts a linear  $\log L_\nu(\rm 2500 \mathring{A}) - \log L_X(\rm 2keV)$ correlation, i.e. the $\mu$ parameter of Equation \ref{equation:model-flavour-1} follows of the form

\begin{align}\label{eq:model-flavour-3}
\mu  =   A \cdot [\log{L_\nu(\rm 2500  \mathring{A}) - 30]} + x_0.
\end{align}

\noindent This is a parametric modelling approach where we drop the UV luminosity bins and fit all QSOs at fixed redshift interval to infer the mean and scatter of the normal distribution of Equation \ref{equation:model-flavour-1}. The total number of free parameters in this case is $3\cdot N_{z-bins}$, i.e. for each redshift bin $A$, $x_0$, and $\sigma$. We adopt the same redshift bins as in the fist model version, i.e. $\Delta z=0.5$. 

 We also consider a variation of the previous model, in which the slope and intercept are linear functions of redshift

\begin{align}\label{eq:model-flavour-4-slope}
A  =   \alpha_{0} + \alpha_s\cdot[(1+z)/(1+z_0)-1],
\end{align}

\begin{align}\label{eq:model-flavour-4-intercept}
x_0  =   \beta_0 + \beta_s\cdot[(1+z)/(1+z_0)-1].
\end{align}

\noindent where the reference redshift, $z_0$,  is set to $1.5$.  The total number of free parameters is 5. All SDSS QSOs in the redshift range $0.5-3.0$ are used in the inference.

The Hamiltonian Markov Chain Monte Carlo code Stan\footnote{\href{https://mc-stan.org}{https://mc-stan.org}} \citep{daniel_lee_2017_1101116} is used to sample the likelihood of Equation \ref{eq:likelihood} and produce parameter posterior distributions for each model flavour described above. For the log-linear $L_X({\rm 2\,keV}) - L_\nu(\rm 2500 \mathring{A})$  correlation of Equation  \ref{eq:model-flavour-3} we also explore if the data prefer redshift-dependence slope and intercept (Equations \ref{eq:model-flavour-4-slope}, \ref{eq:model-flavour-4-intercept}) by performing Bayesian model comparison. For this particular application we also use nested sampling as implemented in the UltraNest\footnote{\url{https://johannesbuchner.github.io/UltraNest/}} package \citep{Buchner2021UN} and based on the MLFriends \citep{Buchner2016UN, Buchner2019UN} Monte Carlo algorithm. Appendix \ref{ap:validation} describes realistic simulations to test and validate the overall hierarchical Bayesian approach in the case of the model parametrization described by Equations \ref{equation:model-flavour-1}, \ref{eq:model-flavour-3}.  

%% file: Analysis_restructured.tex
\section{Results}\label{results}

\begin{figure*}
    \centering
    \includegraphics[scale=0.34]{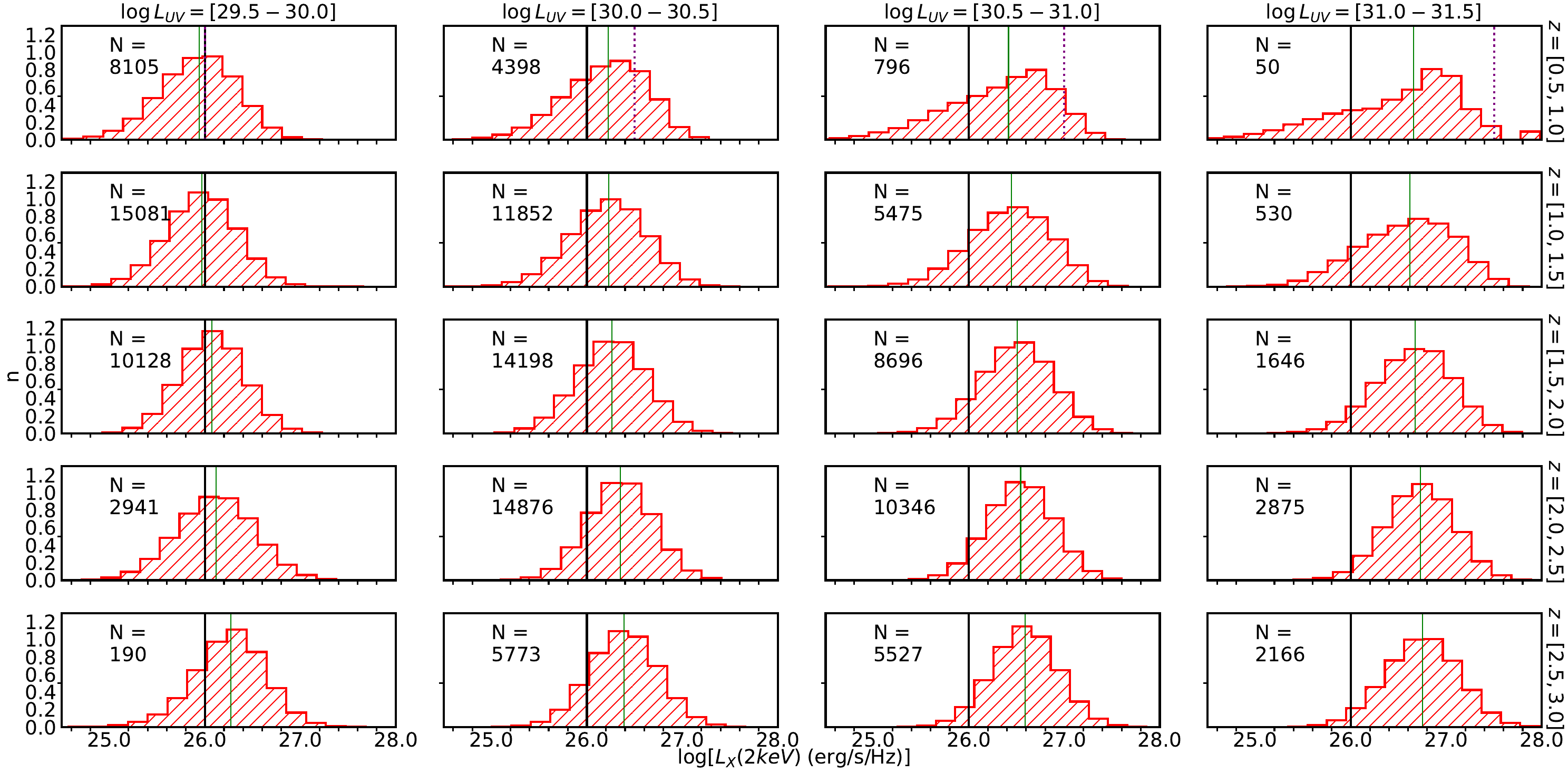}
    \caption{Normalised posterior distribution of $\log{L_X(\rm 2\,keV)}$ in bins of $\log{L_\nu(\rm 2500 \mathring{A})}$ (increasing from left to right) and redshift, $z$ (increasing from top to bottom). The green, solid vertical line shows the median of each distribution. The vertical black line corresponds to $\log{L_X(\rm 2\,keV)} =26$ and is plotted to facilitate comparisons between different panels. The number, $\rm N$, of DRQ16 QSOS in each subsample is shown in the corresponding panel. The dotted line in the first row corresponds to the expected position of the peak of the $\log{L_X(\rm 2\,keV)}$ distribution with increasing $\log{L_\nu(\rm 2500 \mathring{A})}$, assuming a linear $\log{L_X(\rm 2\,keV)}-\log{L_\nu(\rm 2500 \mathring{A})}$ correlation.  }\label{fig:LUV_z_bins}
\end{figure*}

\subsection{Exploring the luminosity and redshift dependence of the X-ray--to--UV correlation}

\begin{figure}
   \centering
   \includegraphics[width=1.0\columnwidth]{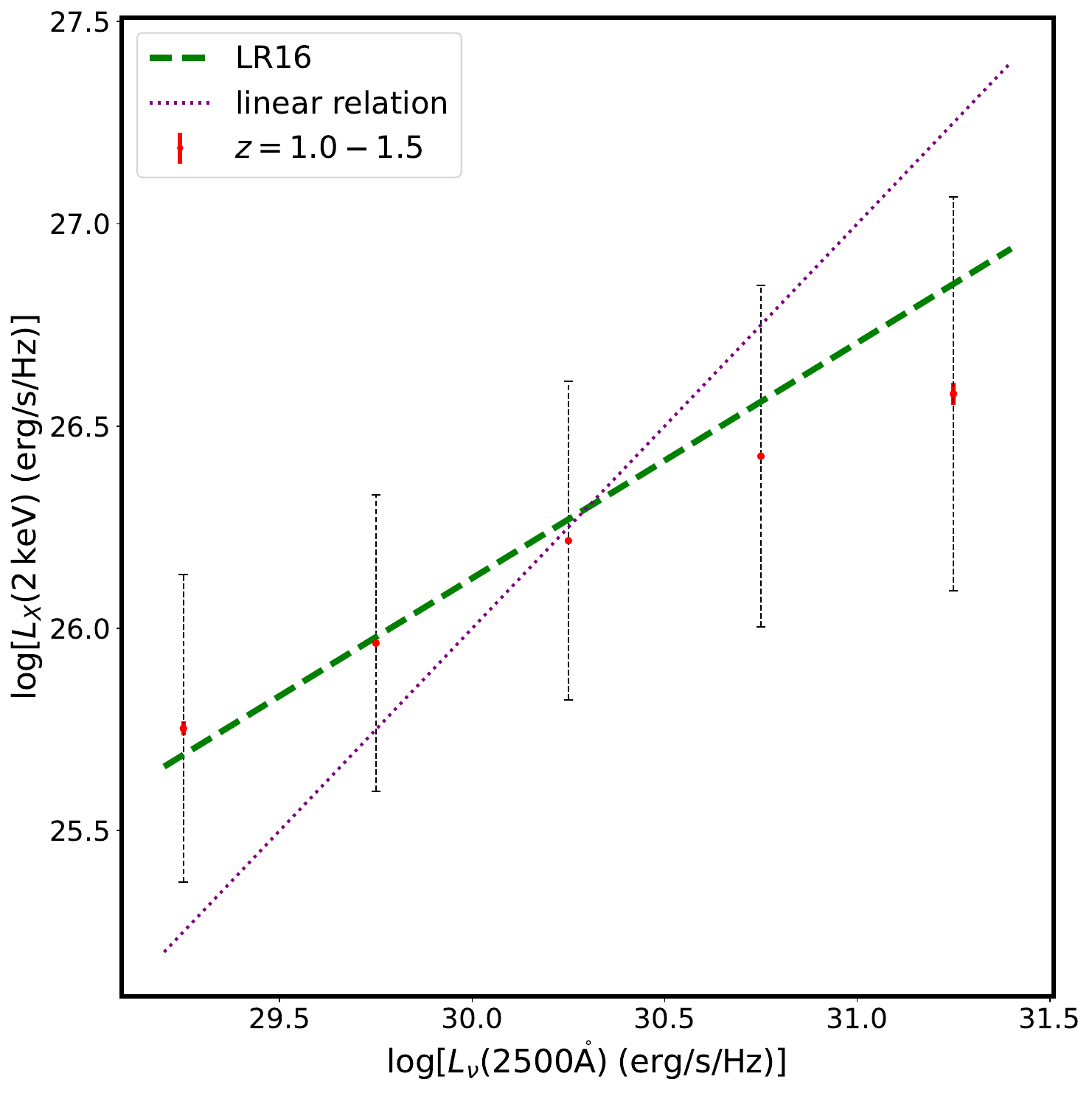}
   \caption{The correlation between the monochromatic luminosities at 2\,keV and 2500\AA. The red data points correspond to the mean of the Gaussian model for the $\log{L_X(\rm 2\,keV)}$ distribution (see Equation \ref{equation:model-flavour-1}) at fixed  $\log{L_\nu(\rm 2500 \mathring{A})}$ luminosity intervals for the redshift bin $z =1.0-1.5$. The x-axis values represent the centers of the $\log{L_\nu(\rm 2500\,\mathring{A})}$ intervals, each with a width of $\Delta \log{L_\nu(\rm 2500\,\mathring{A})} = 0.5$. The red, solid errorbars, correspond to the $1\sigma$ statistical error while, the black, dashed errorbars to the inferred intrinsic Gaussian scatter. The relation of \citet{LR2016} is plotted as a green, dashed line and is consistent with the distribution of our data albeit somewhat steeper. The purple, dotted line shows the linear relation.}\label{fig:lgLx0}
   \end{figure}

\begin{figure}
   \centering
   \includegraphics[width=1.0\columnwidth]{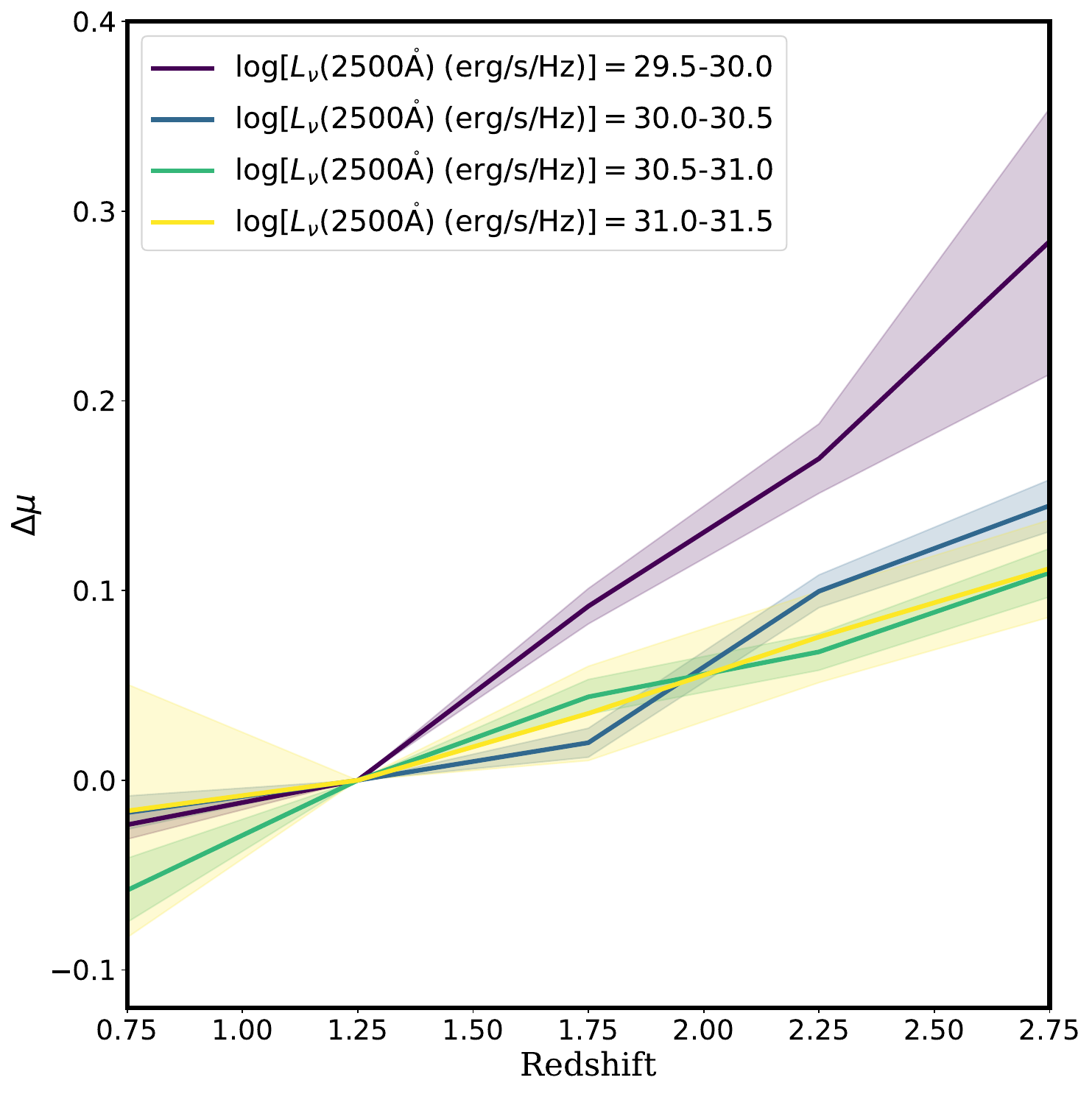}
   \caption{Constraints on the redshift dependence of the monochromatic luminosity $\log{L_X(\rm 2\,keV)}$ distribution, as quantified by the parameter $\Delta \mu (z)$ (Equation \ref{equation:model-flavour-2}) measured with respect to the reference redshift bin, $z =1.0-1.5$. The inferred median $\Delta \mu (z)$ values for each redshift interval are shown with solid lines. The shaded regions correspond to the $1\sigma$ uncertainties. The different colors represent different $\log{L_\nu(\rm 2500 \mathring{A})}$ bins as denoted in the key. The $z - \log{L_\nu(\rm 2500 \mathring{A})}$ bins are the same as in Figure~\ref{fig:LUV_z_bins}. }\label{fig:shift}
   \end{figure}

We first explore in a non-parametric manner the X-ray--to--UV correlation by grouping the data in bins of UV luminosity, $\Delta\log{L_\nu(\rm 2500 \mathring{A})} = 0.5\, \text{dex}$,  and redshift, $\Delta z = 0.5$. For this exercise, we assume the Gaussian model of Equation~\ref{equation:model-flavour-1} for $\log{L_X(\rm 2\,keV)}$ without enforcing an explicit parametric correlation with $\log{L_\nu(\rm 2500 \mathring{A})}$. Figure~\ref{fig:LUV_z_bins} shows the inferred posterior distributions of $\log{L_X(\rm 2\,keV)}$ within different UV luminosity and redshift bins. It allows us to explore non-parametrically how the X-ray luminosity varies with $\log{L_\nu(\rm 2500 \mathring{A})}$ and $z$. The systematic shifts of the peak of the posterior distribution among the panels in a given row of Figure~\ref{fig:LUV_z_bins},  i.e. at fixed redshift, indicate a luminosity dependence. Similarly, changes in the distributions along a column of panels  of Figure~\ref{fig:LUV_z_bins},  i.e., within a fixed $\log{L_\nu(\rm 2500 \mathring{A})}$ interval, can be interpreted as redshift evolution.

We first turn to the $\log{L_\nu(\mathrm{2500\,\mathring{A}})}$ dependence. Figure~\ref{fig:LUV_z_bins} shows a clear systematic increase of the peak of the posterior toward higher UV luminosities for panels at fixed redshift. The emerging trend is sublinear, as highlighted by comparison with the purple, dotted line, shown at the first row and corresponding to a shift of the peak by $\Delta\log{L_X(\rm 2\,keV)} = 0.5$, i.e., equal to the increase in $\log{L_\nu(\mathrm{2500\,\mathring{A}})}$ in each column. This is further demonstrated in Figure~\ref{fig:lgLx0}, which plots  for the redshift bin $z=1.0-1.5$ the median inferred $\log{L_X(\rm 2\,keV)}$ for the different UV luminosity intervals shown in  Figure~\ref{fig:LUV_z_bins}. The correlation recovered by this exercise is in reasonable agreement with what is found in previous studies. As an example, the relation of \citet{LR2016} is plotted for comparison. A linear relation is also shown in the same figure to highlight the sublinear nature of the observed correlation.
   
Another interesting feature of Figure~\ref{fig:LUV_z_bins} is the decrease of the broadness of the plotted posterior distributions with redshift, as well as some skewness of the lower-redshift distributions towards low X-ray luminosities. These tails are more prominent for the panels that correspond to low-redshifts and high-UV luminosities. In this regime, the X-ray observations have sufficient depth to provide strong constraints on the inferred X-ray luminosities of individual sources, despite the imposed prior Gaussian distribution for the overall population. These tails disappear at higher redshifts and the distributions become narrower probably due to the sensitivity of the X-ray observations. In that respect, deeper X-ray observations, e.g. using the full depth of the coadded  eROSITA surveys (eRASS1 to eRASS5), could help further explore the tails of the distributions in  Figure ~\ref{fig:LUV_z_bins}. The subset of X-ray faint QSOs \citep{Pu2020,Zhang2023,Trefoloni2024} for their UV luminosity will be further discussed in Section \ref{sec:discussion}.

Next, we explore the redshift evolution of the $\log{L_\nu(\rm 2500 \mathring{A})}-\log{L_X(\rm 2\,keV)}$ correlation. Comparison of the X-ray luminosity posterior distributions along the redshift columns of Figure~\ref{fig:LUV_z_bins} suggests systematic variations of their peaks. For QSOs within the two least UV luminous bins, the mode of the inferred $\log{L_X(\rm 2\,keV)}$ posterior distribution shifts to higher values with increasing redshift. This trend becomes weaker for QSOs in the two most UV-luminous intervals shown in Figure~\ref{fig:LUV_z_bins}. We further explore this by adopting the model of Equation~\ref{equation:model-flavour-2} to investigate the systematic shifts of the X-ray luminosity posterior distributions  of Figure~\ref{fig:LUV_z_bins}. In more detail, we retain the same grouping in $\log{L_\nu(\rm 2500 \mathring{A})} - z$ bins and assume that, for fixed $\log{L_\nu(\rm 2500 \mathring{A})}$, the mean of the Gaussian distribution that describes the 2\,keV luminosity (Eq.~\ref{eq:model-gh}) at a given redshift deviates by a shift $\Delta\mu$ relative to the mean of a reference redshift bin (assumed to be the one at $z_{0}=1.0-1.5$; Eq.~\ref{equation:model-flavour-2}). 
The results of this analysis are presented in Figure~\ref{fig:shift}. Each line shows the derived shift $\Delta\mu$ as a function of redshift and for each of the four $\log{L_\nu(\rm 2500 \mathring{A})}$ intervals. The shaded regions correspond to the $1\sigma$ uncertainties. Relative to the reference redshift bin ($1.0 <z < 1.5$), the shift $\Delta\mu$  increases toward earlier epochs. However, the amplitude of that trend is UV-luminosity dependent, i.e. decreases  with increasing $L_\nu(\rm 2500 \mathring{A})$. Our results are consistent with a faster evolution of the two least UV-luminous bins, supporting the findings based on the visual inspection of Figure~\ref{fig:LUV_z_bins}.

The results based on the two previous models  indicate an evolution of the X-ray/UV correlation, which seems to be originating from the faster redshift evolution of the least UV-luminous bins in Figure \ref{fig:LUV_z_bins}. The models adopted so far do not impose a specific parametrization linking the UV and X-ray luminosity. 

Next, we explore the parametric model given by Equation~\ref{eq:model-flavour-3} and described in detail in Section~\ref{sec_Method}. To this end, the sample is divided into distinct redshift intervals. Within each of them, the whole dynamic range of $\log{L_\nu(\rm 2500 \mathring{A})}$ is fit to infer the posterior distributions of $\log{L_X(\rm 2\,keV)}$ of individual QSOs, as well as the posterior distributions of the parameters at the population level , i.e., the slope, normalization and scatter, $A,x_0 \text{ and } \sigma$, respectively, of Equation~\ref{eq:model-flavour-3}.

\begin{figure}
   \centering
   \includegraphics[width=1.0\columnwidth]{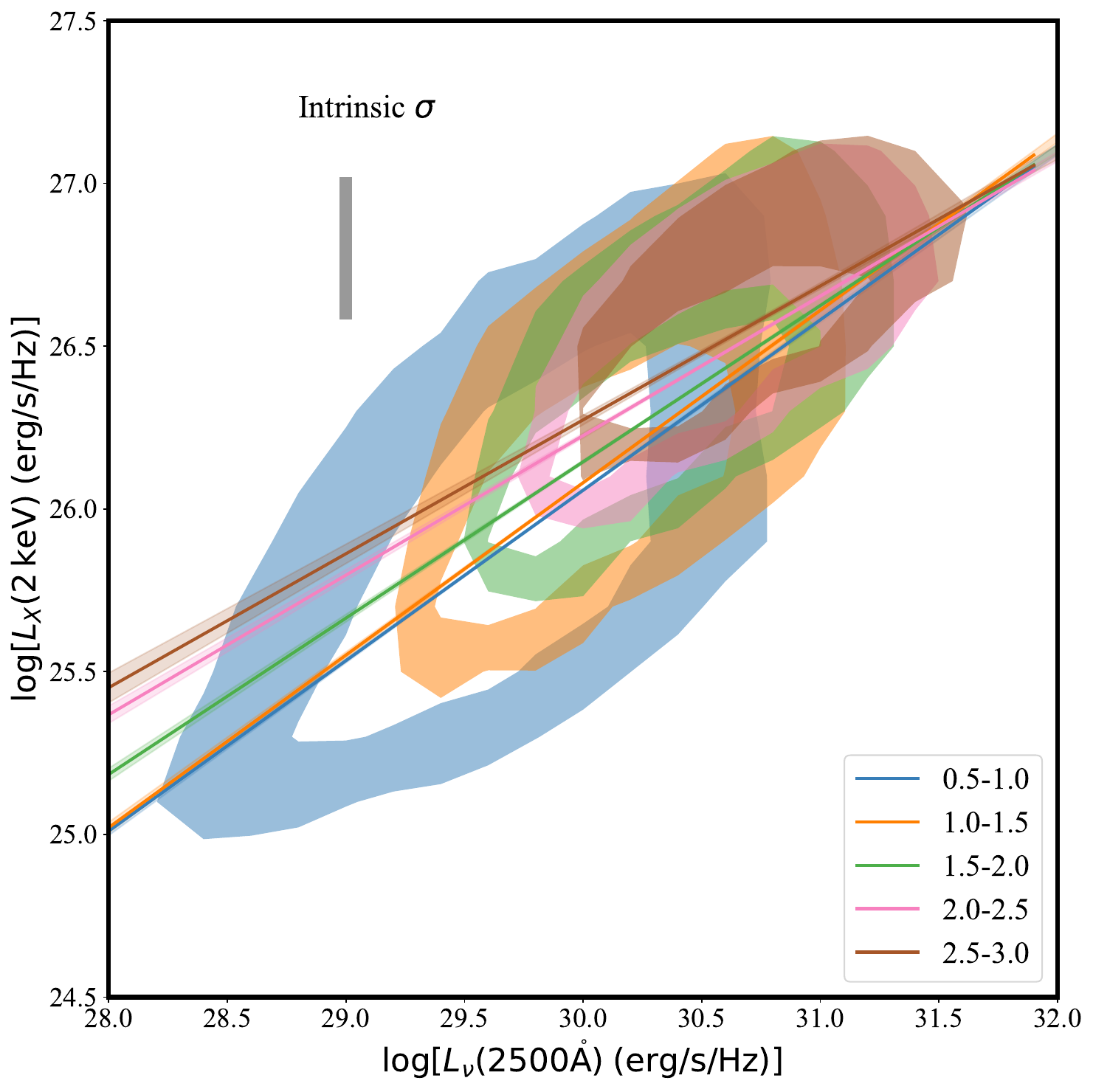}
   \caption{ Inferred $L_X({\rm 2\,keV}) - L_\nu(\rm 2500 \mathring{A})$ correlation  assuming the log-linear model of Equation \protect\ref{eq:model-flavour-3}. The lines correspond to the median of the posterior distributions for the slope ($A$) and intercept ($x_0$), also shown in Table \ref{tab:parameters}. The different colours correspond to the redshift intervals indicated in the legend. The light shaded regions associated with each line correspond to the 68th ($1\sigma$) confidence interval around the median. The complex convex shaded regions are density contours that correspond to the superposition of the posterior distributions of individual QSOs within a given redshift bin. The inner and outer boundaries of each convex shaded region correspond to the iso-density contours that enclose, respectively,  $1\sigma$ (68\%) and $2\sigma$ (95\%) of the posterior at fixed UV luminosity. The inner area of these regions, i.e. the one enclosed by the $1\sigma$ iso-density contour, is left empty for clarity and to facilitate the visualization. A typical value of the intrinsic scatter ($\sigma \approx0.4$\,dex, see Fig.~\ref{fig:params_z}) around the log-linear correlation of Equation \protect\ref{eq:model-flavour-3} is shown with the gray shaded vertical line on the top left.} \label{fig:Stan_lines_z}%
    \end{figure}

\begin{figure}
   \centering
   \includegraphics[scale=0.5]{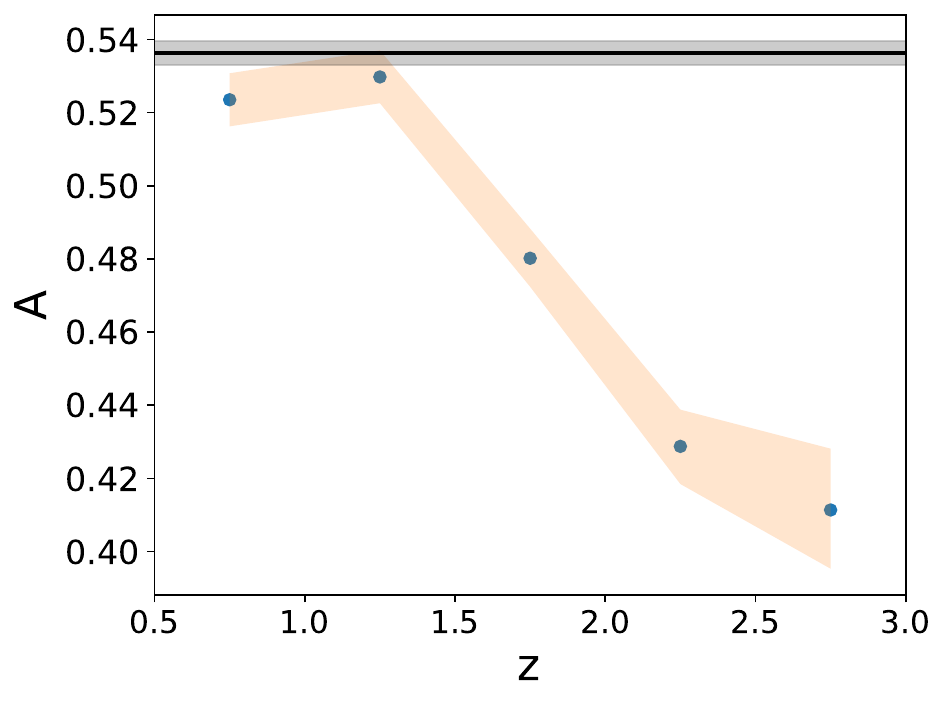}
   \includegraphics[scale=0.5]{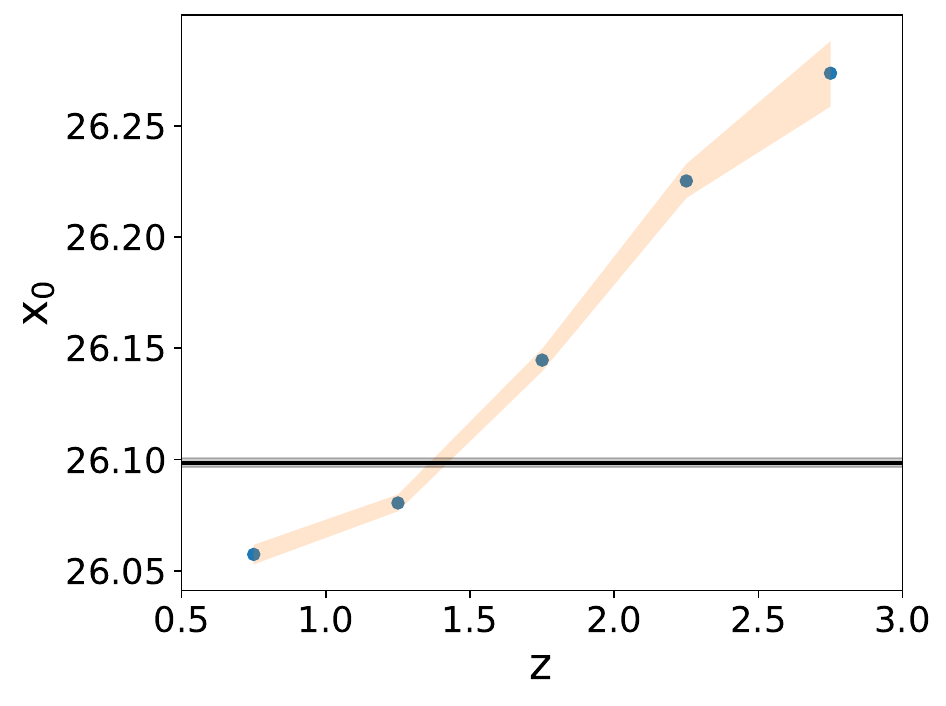}
   \includegraphics[scale=0.5]{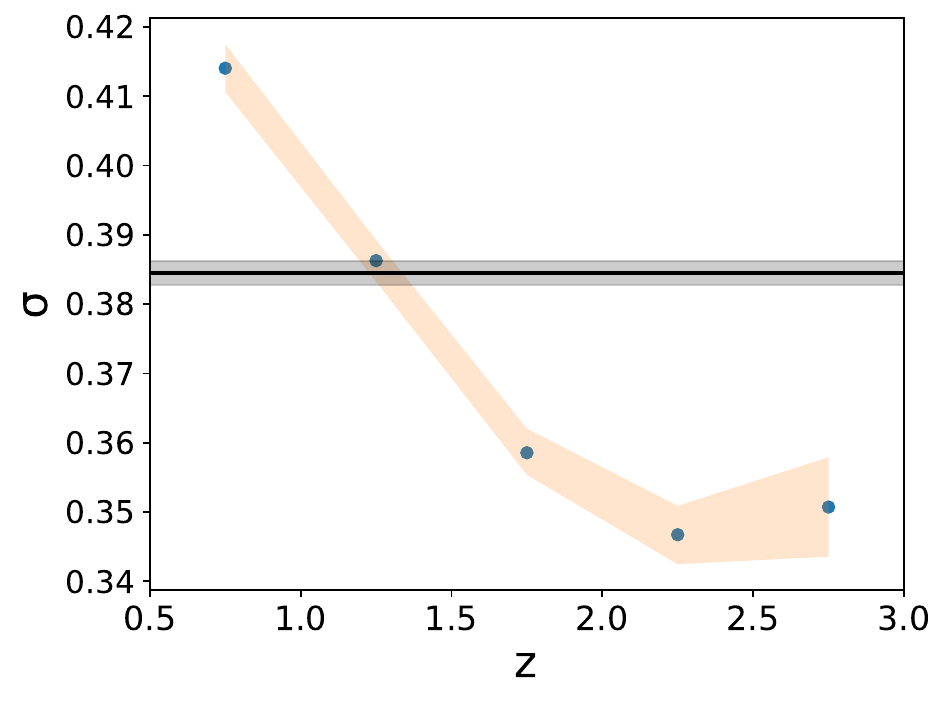}  
   \caption{Evolution as a function of redshift of the individual parameters of the log-linear $L_X({\rm 2\,keV}) - L_\nu(\rm 2500 \mathring{A})$ model, $A$ (top panel), $x_0$ (middle panel)  and  $\sigma$ (bottom panel). The blue data points correspond to the median of the parameter distributions at each of the redshift intervals presented in Table \protect\ref{tab:parameters}. The orange shaded region denote the $1\sigma$ confidence intervals of the parameter posteriors. The inferred parameters in the case of the fit to the full redshift range of the DRQ16 sample (assuming no redshift dependence) are shown in each panel with the horizontal black line (median of posteriors) and gray shaded regions ($1\sigma$ uncertainties).
   }\label{fig:params_z}%
\end{figure}

The inferred correlations and their uncertainties are illustrated in Figure~\ref{fig:Stan_lines_z}. These results suggest a mild but statistically significant\footnote{Using {\sc ultranest} we find that the ratio of the Bayesian evidence (Bayes factor) between the redshift-dependent parameterisation of Equations \ref{eq:model-flavour-4-slope}, \ref{eq:model-flavour-4-intercept} (evidence $Z_{evol}$) and the linear X-ray/UV correlation model with no dependence of the slope/intercept on redshift (e.g. Equation \ref{eq:model-flavour-3}, evidence $Z_{noevol}$) is $\log_{10}(Z_{evol}/Z_{noevol})\approx93$. According to the Jeffrey's scale this is decisive evidence in favour of redshift evolution.} and systematic evolution of the correlation, with the general trend being a decreasing X-ray luminosity at lower redshifts. Notably, the evolution within the explored redshift range is more pronounced toward lower UV luminosities. At higher UV-luminosities, the separation between the lines diminishes. Such a UV luminosity-dependent evolution manifests itself as a change of the slope of the adopted log-linear model, i.e. a flattening to higher redshift. 
Also shown in Figure~\ref{fig:Stan_lines_z} are the density contours of the posteriors of individual QSOs within each redshift interval. The inferred values of the log-linear model parameters at different redshift intervals are shown in Table \ref{tab:parameters} and are plotted as a function of redshift in Figure~\ref{fig:params_z}. 

We further explore the redshift evolution of the $L_X({\rm 2\,keV}) - L_\nu(\rm 2500 \mathring{A})$ correlation using the parametric model described by Equations \ref{eq:model-flavour-4-slope}, \ref{eq:model-flavour-4-intercept}. For this exercise the full sample in the redshift interval $z=0.5-3$ is used. Bayesian model comparison \citep[UltraNest,][]{, Buchner2021UN} shows that the a redshift dependent intercept and slope of the $L_X({\rm 2\,keV}) - L_\nu(\rm 2500 \mathring{A})$ correlation is significantly preferred over models in which either one or both of these parameters are fixed with redshift. Table \ref{tab:parameters-full-parametric} shows the inferred parameters for this  model flavour. The model is graphically shown in the Appendix Figure \ref{fig:Stan_zevol}. 

We caution that the SDSS DRQ16 QSO sample is limited by optical flux, which in effect translates to a narrower UV-luminosity baseline with increasing redshift. The validation simulations presented in Appendix \ref{ap:validation} show that this effect is unlikely to be responsible for the inferred evolution pattern of the X-ray/UV evolution, i.e. a flattening toward higher redshift.

\subsection{Comparison to literature results}

The $L_X({\rm 2\,keV}) - L_\nu(\rm 2500 \mathring{A})$ correlation has been extensively studied mainly as a reflection of the energetic interplay between the dominant components of the accretion flow in AGN. Although there is a rich body of literature on this topic, most previous studies are based on relatively limited samples. Several key aspects of this work are different. For the first time we utilize all-sky X-ray data and apply a novel, robust methodology that maximizes the information extracted from the observations. It is thus interesting to investigate the level of agreement between this work and the findings of previous authors both as a sanity check for our work but also to understand if and how the aforementioned advancements update our knowledge on the X-ray--to--UV correlation.

\begin{figure}
   \centering
   \includegraphics[width=1.0\columnwidth]{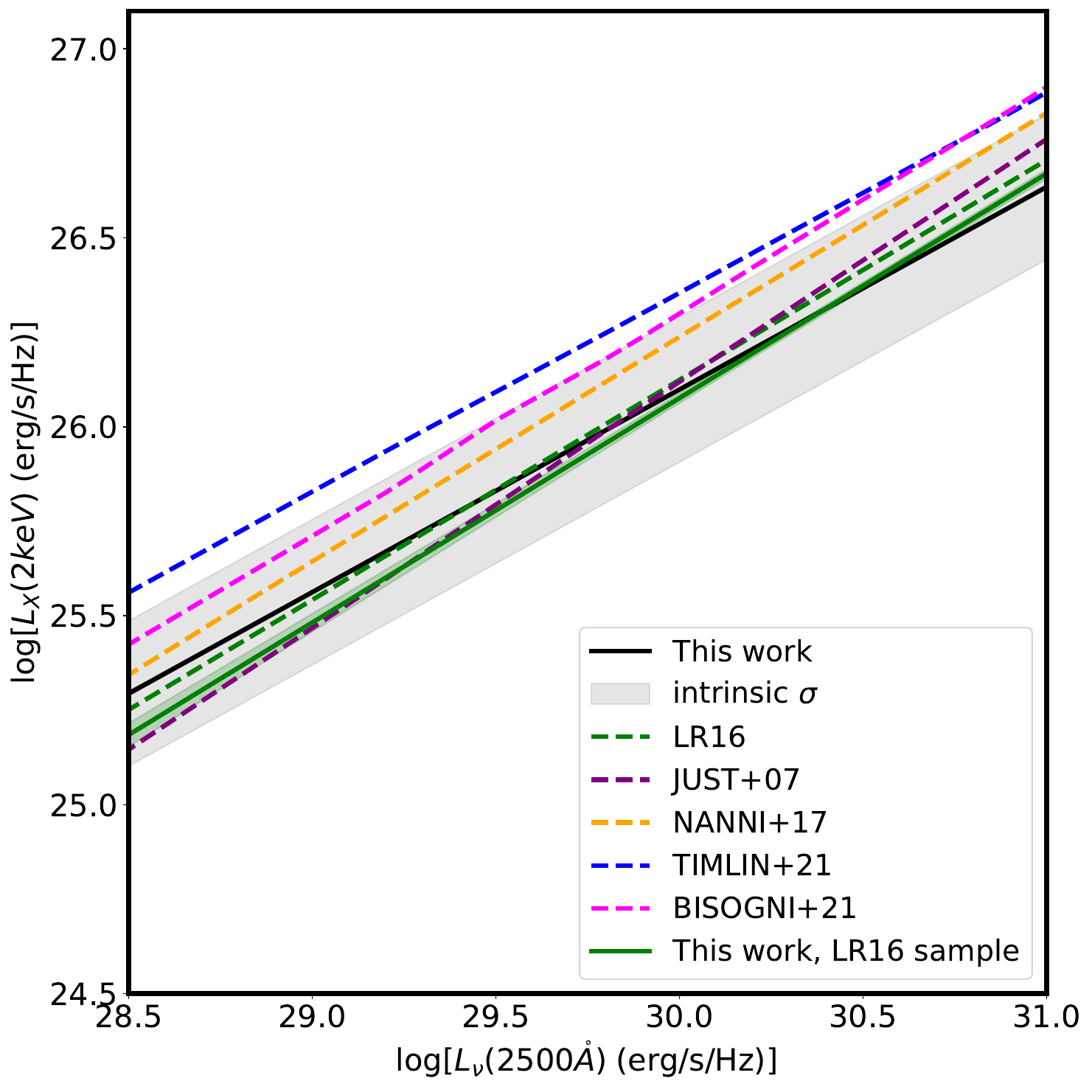}
   \caption{Comparison of our fit (black solid line) to the full DRQ16 QSO sample ($z=0.5-3.0$) with best fit $\log{L_\nu(\rm 2500 \mathring{A})}-\log{L_X(\rm 2\,keV)}$ relations from the literature, as denoted in the key. The best fit relation retrieved by application of our methodology on the sample of LR16 is shown with the solid green line. The shaded gray area is the intrinsic scatter, $\sigma$, inferred for our total sample.} \label{fig:comp_literature}
   \end{figure}

We start by applying the log-linear model of Equation \ref{eq:model-flavour-3} to QSOs over the full redshift interval, $z = 0.5-3$. This approach ignores any redshift dependence of the model parameters but enables the comparison with literature results that do not consider possible redshift evolution effects. The recovered log-linear model parameters for our full QSO sample are shown in Table \ref{tab:parameters} (first row).  Figure~\ref{fig:comp_literature} compares our recovered correlation with previous results from \cite{JUST2007}, \cite{LR2016}, \cite{Nanni2017}, \cite{Bisogni2021}, \cite{Timlin2021}. Discrepancies are evident between the fit obtained from the current work and some of the previous studies. Our inferred correlation is broadly consistent with those reported by \citet{JUST2007}, \citet{LR2016} and \citet{Nanni2017}. The largest discrepancy ($\approx 0.5$\,dex) is found  for \citet{Timlin2021}.

It is important to note that these earlier results are based on samples much smaller than ours (up to three orders of magnitude), often limited to X-ray detections, and/or fulfilling strict selection criteria (e.g. X-ray signal-to-noise ratio, UV spectral slope). For example, \citet{JUST2007} select 59 of the most optically luminous, radio-quiet, and non-BAL quasars from SDSS and high-redshift surveys using a combination of \textit{Chandra}, \textit{XMM-Newton}, and \textit{ROSAT} data. Their analysis includes both X-ray detections and upper limits. \citet{LR2016} cross-match SDSS quasars with \textit{XMM-Newton} and \textit{Chandra} and apply stringent quality cuts to remove radio-loud sources, BALs, and quasars with low X-ray photon indices or significant absorption.  Their final sample numbers 808 QSOs and includes X-ray upper limits. \citet{Bisogni2021} further refined the above selection by fitting the optical–UV SED to exclude dust-reddened or host-dominated sources, obtaining a sample of 1,142 bright, blue quasars, but included only sources with reliable X-ray detections. \citet{Timlin2021} applied similar criteria to SDSS and X-ray data, focusing on high-quality multiwavelength coverage and excluding contaminated sources, yielding a sample of 482 quasars based solely on X-ray detections. \citet{Nanni2017} targeted high-redshift ($z \gtrsim 4$) quasars with available X-ray observations, excluding radio-loud and BAL quasars and ensuring rest-frame UV coverage. They included both detections and upper limits in their analysis of 53 sources.

It is not straightforward to replicate the selection criteria above to explore the origin of the discrepancy between our analysis and previous studies. This is nevertheless possible in the case of \citet{LR2016} that provide sufficient information on their final QSO catalogue. We select from our sample the same QSOs as those used by \citet{LR2016} and apply our methodology to this smaller subset. The results of this exercise are shown with the green solid line in Figure~\ref{fig:comp_literature}. The level of discrepancy is reduced to the level of statistical uncertainties. This emphasizes the importance of QSO sample selection and/or the methodology when comparing the UV/X-ray correlation among different studies.

\begin{table}
    \caption{Median and $1\sigma$ uncertainties derived from the posterior distributions of the parameters \( A \), \( x_0 \), and \( \sigma \) that describe the log-linear relation of Equation \ref{eq:model-flavour-3}. Shown are the total sample, spanning the full redshift range $z=0.5-3$, as well as the QSO sub-samples selected in narrower redshift intervals.}
    \label{tab:parameters}
    \centering
    \begin{tabular}{l c c c}
        \toprule
        z & \( A \) & \( x_0 \) & \( \sigma \) \\
        \midrule
        0.5 -- 3.0  & \( 0.537 \pm 0.003 \ \) & \( 26.099 \pm 0.002 \) & \( 0.385 \pm 0.002 \) \\
        0.5 -- 1.0  & \( 0.554 \pm 0.008 \) & \( 26.057 \pm 0.004  \) & \( 0.414 \pm 0.004 \) \\
        1.0 -- 1.5  & \(0.530 \pm 0.007 \) & \( 26.081 \pm 0.004 \) & \( 0.386 \pm 0.003 \) \\
        1.5 -- 2.0  & \( 0.480 \pm 0.008 \) & \( 26.145 \pm 0.005 \) & \( 0.359 \pm 0.003 \) \\
        2.0 -- 2.5  & \(0.429 \pm 0.010 \) & \( 26.225 \pm 0.007 \) & \( 0.348 \pm 0.003 \) \\
        2.5 -- 3.0  & \(0.411 \pm 0.017 \) & \( 26.274 \pm 0.014 \) & \( 0.351 \pm 0.006 \) \\
        \bottomrule
    \end{tabular}
\end{table}

\begin{table}
    \caption{Median and $1\sigma$ uncertainties derived from the posterior distributions of the parameters \( \alpha_0 \), \( \alpha_s \) of Eq. \ref{eq:model-flavour-4-slope}, \( \beta_0 \), \( \beta_s \) of Eq. \ref{eq:model-flavour-4-intercept} and \( \sigma \).}
    \label{tab:parameters-full-parametric}
    \centering
    \begin{tabular}{c r}
    \toprule
        $\alpha_0$ & $0.485\pm0.004$\\
        $\alpha_s$ & $-0.08\pm0.01$ \\
        $\beta_0$ & $26.118\pm0.003$ \\
        $\beta_s$ & $0.20\pm0.01$  \\
        \( \sigma \) & $0.380\pm0.002$\\
        \bottomrule
    \end{tabular}
\end{table}

The aforementioned authors do not support a redshift dependence of the X-ray--to--UV correlation. However, indications for such an evolution of the correlation with redshift has been reported in the literature \citep[e.g.,][]{Wang2022_copulas, Khadka2021b, Khadka2022, Rankine2024}. Among these works, the three former approach the correlation from a cosmological point of view, while \citet{Rankine2024} focus on a detailed and careful modeling of its redshift evolution using a Bayesian methodology similar to ours that also accounts for non X-ray detections. Their similar point of view allows a direct quantitative comparison with our findings. Specifically, we 
 focus our comparison on their Figure 7, which uses a slightly different parameter space, $\log{L_\nu(\rm 2500 \mathring{A})} - \log{L_X(\rm{0.5-10 \,keV})}$. The conversion between the monochromatic $L_X(\rm 2\,keV)$ and $L_X(\rm{0.5-10 \,keV})$ requires assumptions on the X-ray spectral model. Consistently with the rest of the analysis we assume a power-law photon-flux spectrum with $\Gamma = 1.9$.





\begin{figure}
    \centering
    \includegraphics[width=1.0\columnwidth]{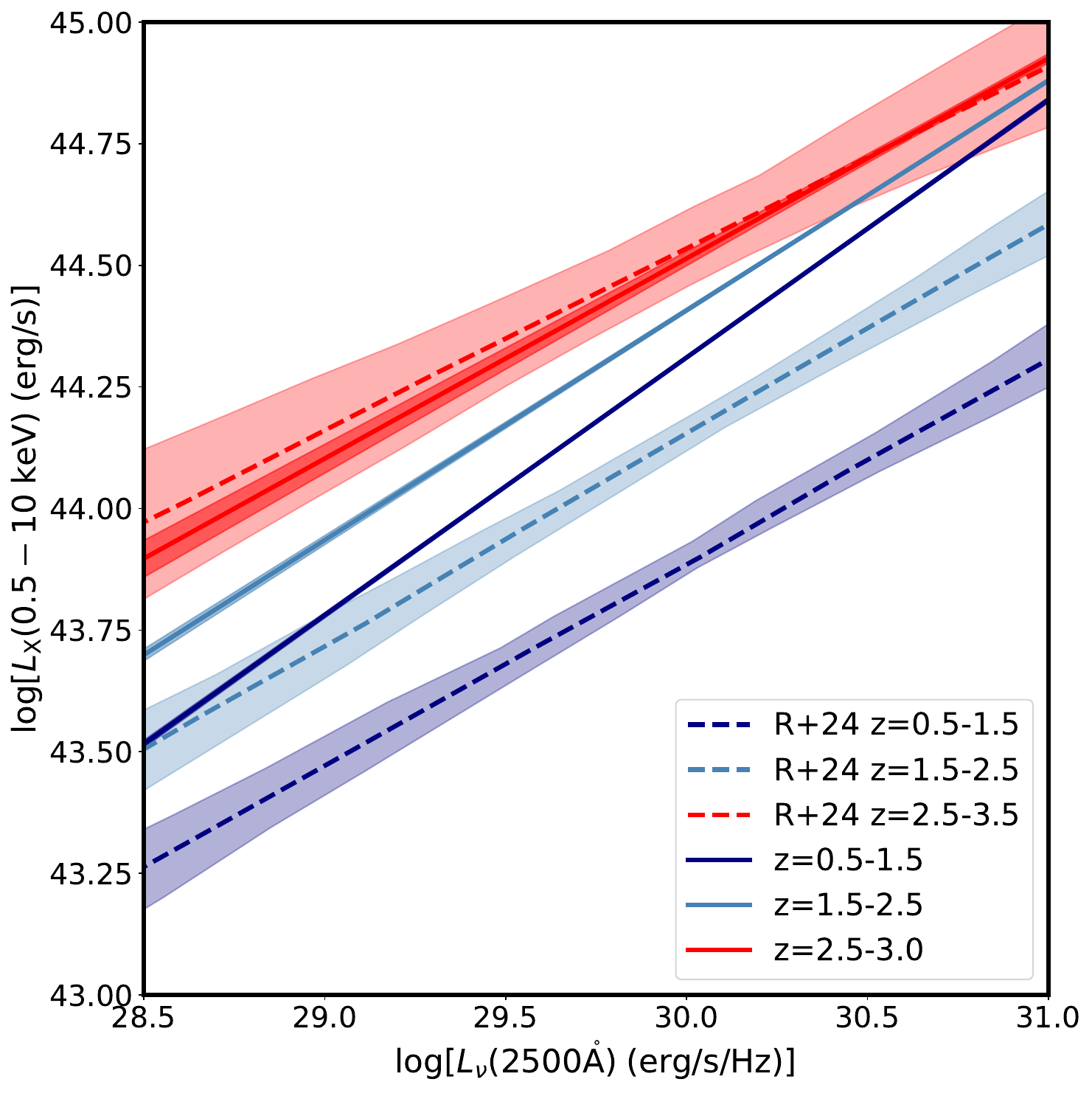}
    \caption{ Inferred $\log{L_\nu{2500\,\mathring{A}}}-\log{L_X(\rm{0.5-10 \,keV})}$ 
    correlation using our sample (solid lines) in comparison with  the model (iii) of \citet[][dashed lines; taken from their Figure 7]{Rankine2024} 
    for three redshift intervals, $z = 0.5-1.5$ (blue), $z= 1.5-2.5$ (light blue), $z =2.5-3.5$ (red). The shaded areas correspond to the%
    $1 \sigma$ statistical uncertainties. 
    }
    \label{fig: Compare_rankine}
\end{figure}

Figure~\ref{fig: Compare_rankine} plots our results (solid lines) along with those of \citet{Rankine2024} (dashed lines) for the same three redshift intervals, $z=0.5-1.5$ (blue), $z=1.5-2.5$ (light blue), $z=2.5-3.5$ (red). We show their results for their model version (iii). Although this it is not their preferred model, it is the one that is closer to our approach and thus best-suited for a comparison. The evolution found is in the same direction for both cases, that is, higher X-ray luminosities at higher redshift. However, the amplitude of the evolution is different. \citet{Rankine2024} find an increase in the normalization of $\Delta\log{L_X(\rm{0.5-10 \,keV})} \approx 0.75 \text{\,dex}$ to $z\approx3.5$ and a constant slope. On the other hand, we recover a flatter slope with increasing redshift and an amplitude of $\approx 0.25 \text{ dex}$ at $\log L_\nu(\rm 2500 \mathring{A}) \approx28.5$ (erg/s), where the evolution is stronger. As an attempt to resolve and comprehend this discrepancy, we apply our methodology on their sample. Specifically, we select the same objects and use both their $L_\nu(\rm 2500 \mathring{A})$ and ours, to find hardly any dependence on redshift in both cases. The origin of the discrepancy with \citet{Rankine2024} is not clear and is likely associated with differences in the adopted methodology.

\subsection{Exploring the Eddington ratio dependence of the X-ray--to--UV correlation}

\begin{figure*}
    \centering
    \includegraphics[scale=0.36]{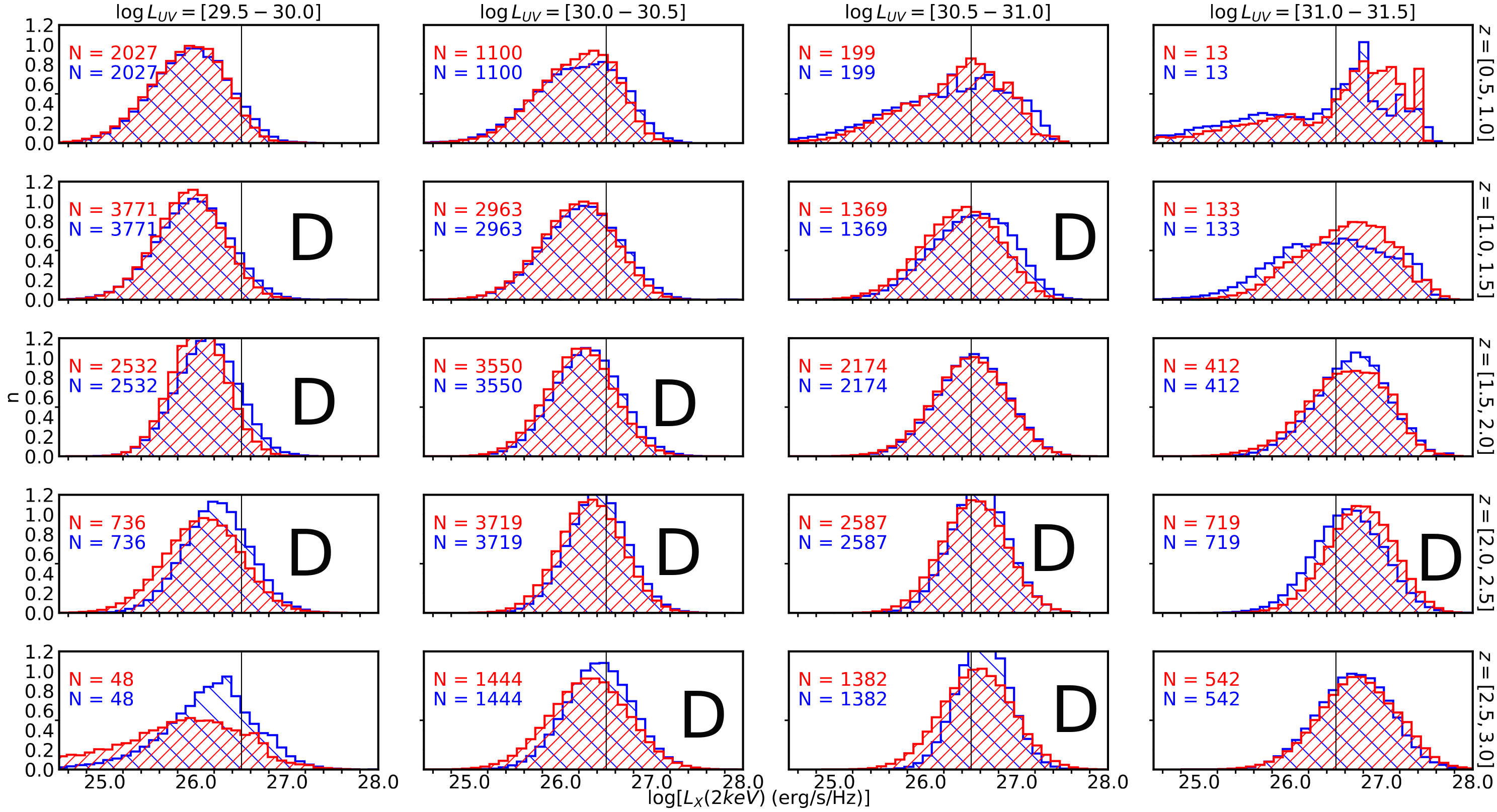}
    \caption{Posterior distribution of $L_X({\rm 2\,keV})$ in bins of $\log{L_\nu(\rm 2500 \mathring{A})}$ (increasing from left to right) and redshift, $z$ (increasing from top to bottom). The blue and red distributions correspond to the $25\%$ lowest and highest Eddington ratio sources of each $\log{L_\nu(\rm 2500 \mathring{A})}-z$ bin. The black capital "D" signifies the bins where the highest- and lowest- Eddington ratio subsamples are drawn from different distributions, based on the result of the Kolmogorov-Smirnov two-sample test  at a $5\%$ confidence level. }
    \label{fig: LUV_z_bins_ledd}
\end{figure*}

\begin{figure*}
    \centering
    \includegraphics[scale=0.3]{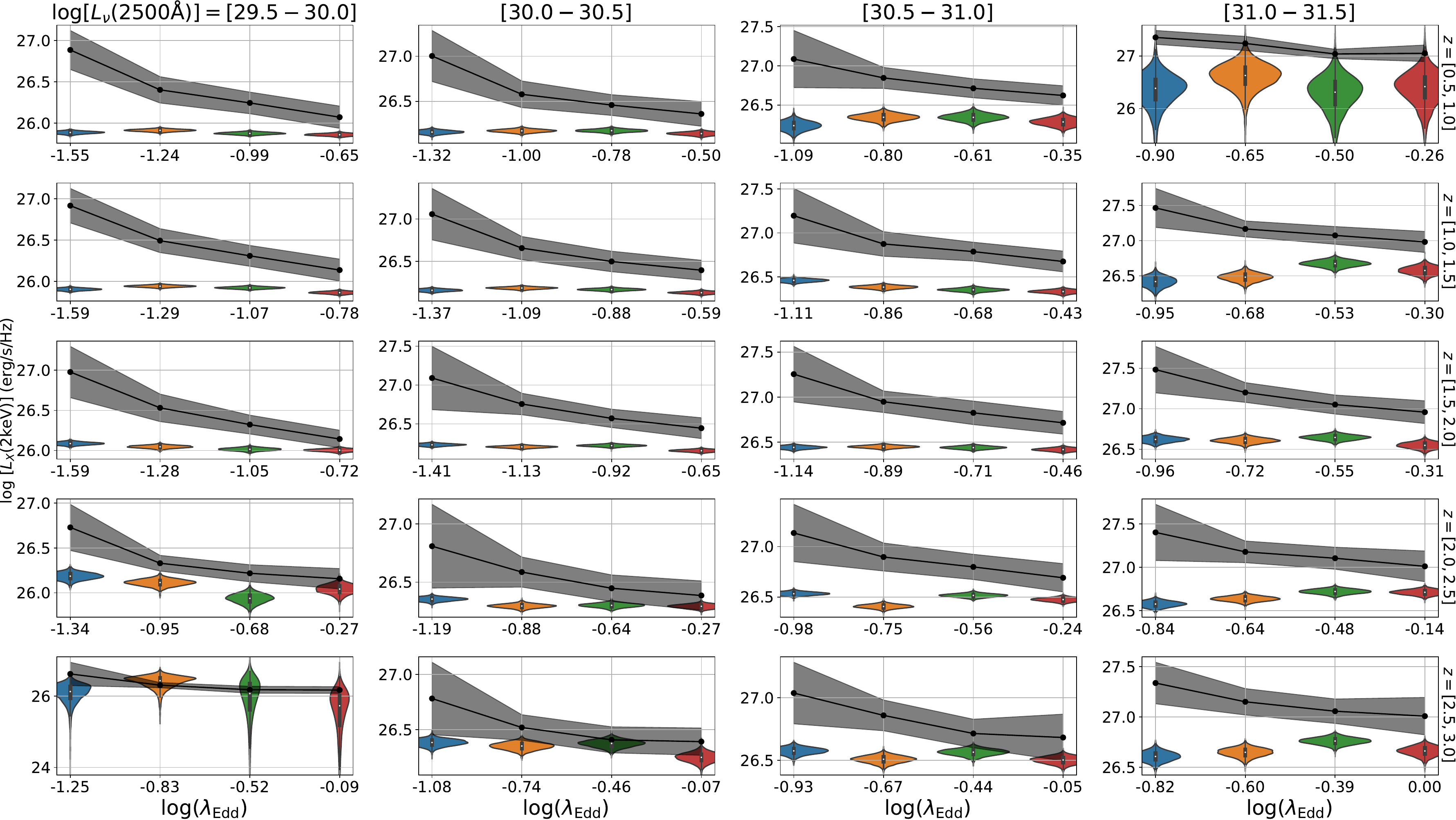}
    \caption{Eddington ratio dependence of the $L_X(\rm 2\,keV)$ inferred from observations (violin symbols) in comparison with the predictions of the \textsc{qsosed} model \protect\citep[][grey shaded regions and black points connected with solid line]{Kubota_Done2019}. Each panel corresponds to DRQ16 QSO subsamples binned in $\log{L_\nu(\rm 2500 \mathring{A})}$ (increasing from left to right) and redshift (increasing from top to bottom). The violin symbols in each panel represent the full posterior distribution of the inferred $L_X(\rm 2\,keV)$ in four Eddington ratio bins. These are defined by splitting the QSO sample at the 25th, 50th and 75th quantiles of the Eddington ratio distribution. The blue, yellow, green and red violin  colours are for the lowest, intermediate low, intermediate high and highest quantiles, respectively. We caution that because of this definition of the Eddington ratio bins, the x--scale and y--scale is different for each panel. The \textsc{qsosed} model predictions are constructed by estimating for each QSO in a given Eddington ratio, $\log{L_\nu(\rm 2500 \mathring{A})}$ and redshift bin the expected $L_X(\rm 2\,keV)$ for its black hole mass and  Eddington ratio assuming zero spin, an electron temperature for the hot Comptonisation component of 100\,keV, a dissipated corona luminosity in Eddington units of 0.02, an electron temperature for the warm Comptonisation component of 0.2\,keV, a spectral index of the warm Comptonisation component of 2.5 and reflection albedo for the reprocessed flux of 0.3. The black symbols connected with a solid line shown the median predicted $L_X(\rm 2\,keV)$ of all AGN in the a given bin. The grey shaded regions show the 68th confidence interval around the median.} 
    \label{fig: LUV_z_bins_ledd_qsosed}
\end{figure*}
 
 Models predict that the structural properties of the accretion flow vary as a function of accretion power \citep[e.g.,][]{Kubota_Done2018}, thereby linking the dimensions of the X-ray corona and accretion disc to the Eddington ratio. This interplay is expected to introduce a dependence of the X-ray--to--UV correlation on $\lambda_{Edd}$. It is therefore interesting to explore this prediction in our sample. 



The size of our sample allows to proceed in such an investigation by isolating sources within relatively narrow intervals in the multidimensional parameter space defined by $\log{L_\nu(\rm 2500 \mathring{A})}$, redshift  and $\lambda_{Edd}$\footnote{
The Eddington ratio is defined as 
$\lambda_{\mathrm{Edd}} = L_{\mathrm{bol}} / L_{\mathrm{Edd}}$, 
where $L_{\mathrm{Edd}} = 1.26 \times 10^{38} (M_{\mathrm{BH}}/M_\odot)\,\mathrm{erg\,s^{-1}}$. 
The black hole masses ($M_{\mathrm{BH}}$) and bolometric luminosities ($L_{\mathrm{bol}}$) are  from \citet{Wu_Shen2022}.  $M_{\mathrm{BH}}$ is estimated using single-epoch spectra based on the width of broad emission lines and the corresponding continuum luminosities. $L_{\mathrm{bol}}$ is derived from the monochromatic luminosities at 1350, 3000 or 5100\AA\, depending on the source's redshift. The corresponding bolometric corrections are estimated from the mean spectral energy distribution of quasars in \cite{Richards2006}.}. For each $\log{L_\nu(\rm 2500 \mathring{A})}- z$ bin of Figure~\ref{fig:LUV_z_bins} we isolate sources within the $25^{\rm th}$ highest and lowest quartiles of the Eddington ratio distribution. The model described by Equation~\ref{equation:model-flavour-1} is then applied to each subsample to infer the $\log{L_X(\rm 2\,keV)}$ posteriors separately for the lowest and highest quartiles in each $\log{L_\nu(\rm 2500 \mathring{A})}- z$ bin. The results are presented in Figure~\ref{fig: LUV_z_bins_ledd} with the red (high Eddington ratio) and blue (low Eddington ratio) histograms. We caution that the boundaries that define the 25th and 75th quintiles of the $\lambda_{Edd}$ distribution are different for each $\log{L_\nu(\rm 2500 \mathring{A})}-z$ subsample and hence, the  histograms in each panel of Figure~\ref{fig: LUV_z_bins_ledd} do not correspond to the same $\lambda_{Edd}$ intervals.

Visual inspection of the blue and red histograms within each panel of Figure~\ref{fig: LUV_z_bins_ledd} shows no strong dependence of $\log{L_X(\rm 2\,keV)}$ on Eddington ratio. However, for some panels a small but systematic shift of the distribution can be seen with the high-$\lambda_{Edd}$ histogram being shifted towards lower $L_X(\rm 2\,keV)$ values. We use the  Two Sample Kolmogorov-Smirnov Test (KS2test\footnote{A non-parametric test that estimates the probability of two samples being drawn from the same parent population.}) to assess if there are statistically significant differences between the histograms of Figure~\ref{fig: LUV_z_bins_ledd}. The two-sample Anderson–Darling test yields results consistent with the Kolmogorov–Smirnov test.

We choose to use the full posteriors in the KS2test comparison to account for the uncertainties associated with the $L_X(\rm 2\,keV)$ distributions in Figure~\ref{fig: LUV_z_bins_ledd}. For a given $\log{L_\nu(\rm 2500 \mathring{A})}/z$ bin we  sample separately the posteriors of the low and high Eddington ratio sources to generate two realisations, one for each subpopulation, with size equal to the number of sources in the original subsamples. We repeat this exercise to produce 100 realisation pairs, which can then be compared using the KS2tests to estimate 100 $p$-values. Since we perform multiple (100) KS2tests for each bin, we need to apply a correction for the Family-Wise Error Rate (FWER\footnote{The probability of making at least one Type I error (i.e. false positives) among all the hypothesis tests when performing multiple tests.}). We apply the Bonferroni correction by dividing the adopted confidence interval, $\alpha_{0} =0.05$, with the number of realisations. The  corrected value of the confidence interval is therefore $\alpha = \frac{\alpha_{0}}{100}$. For each $\log{L_\nu(\rm 2500 \mathring{A})}-z$ we compare the median of the 100 $p$-values to $\alpha$. If the median value $p_{\rm med} < \alpha $ we conclude that the two distributions are different (this result is denoted as D).  Else the null hypothesis that the two samples are drawn from the same populations (result denoted as S) can not be rejected. The KS2test results for each $\log{L_\nu(\rm 2500 \mathring{A})}-z$ bin are shown on each panel of Figure~\ref{fig: LUV_z_bins_ledd} with a capital "D" letter signifying that the two subsamples are drawn from different distributions. This analysis indicates that the $\log{L_X(\rm 2\,keV)}$ distributions of the low/high Eddington ratio sources are different at a confidence level of $5\%$ for $50\%$ of the panels shown in Figure~\ref{fig: LUV_z_bins_ledd}. There is no clear systematic trend with $L_\nu(\rm 2500 \mathring{A})$ or $z$.

The weak, if any at all,  dependence of the inferred X-ray--to--UV correlation to the Eddington ratio motivates us to explore the predictions of accretion flow models to assess the amplitude of the expected differences. For this exercise, we choose to use the models of \cite{Kubota_Done2018}. The \textsc{QSOSED} model assumes a radially stratified accretion flow composed of three energetically coupled zones: an outer, optically thick Novikov--Thorne disc emitting thermal UV--optical radiation; an intermediate warm Comptonizing region producing the soft X-ray excess; and an inner hot corona responsible for the hard X-ray power-law tail. The model links these components self-consistently through the local energy budget, fixing the coronal size and spectral indices according to the fraction of accretion power dissipated in each zone. It has been shown to reproduce reasonably well the observed optical-to-X-ray spectral energy distributions of luminous AGN and their trends with Eddington ratio \citep[e.g.][]{Kubota_Done2018,Mitchell2023,Temple2023}, and the warm-Comptonization component adopted in the model is supported by large-sample analyses of AGN soft-excess spectra \citep[e.g.][]{Petrucci2018}. However, for a discussion of the model limitations and caveats, such as parameter degeneracies, calibration on a small AGN sample, and possible deviations from a steady thin-disc geometry, see \citet{Kubota_Done2018}, \citet{Kynoch2023}, and \citet{Temple2023}.

Figure \ref{fig: LUV_z_bins_ledd_qsosed} plots the inferred X-ray luminosity as a function of Eddington ratio for subsamples selected within the same redshift and $\log{L_\nu(\rm 2500 \mathring{A})}$  intervals as in Figure  \ref{fig: LUV_z_bins_ledd}. Each redshift and $2500\mathring{A}$, monochromatic luminosity subsample is split into four equal size groups that are separated by the 25th, 50th and 75th  quantiles of the Eddington ratio distribution of the subsample under consideration. As a result the adopted Eddington ratio bins are different for each panel of Figure \ref{fig: LUV_z_bins_ledd_qsosed} and depend on the $\lambda_{Edd}$ distribution of each $z\,/\,\log{L_\nu(\rm 2500 \mathring{A})}$  subsample. The x--axis range in each panel of Figure \ref{fig: LUV_z_bins_ledd_qsosed}  is therefore different. The observationally inferred $L_X(\rm 2\,keV)$  posterior distributions are represented by the violin symbols in each panel. These are further compared to the predictions of the \textsc{qsosed} model of \cite{Kubota_Done2019} shown with the shaded regions. These are estimated as follows. For each QSO with an estimated black hole mass and Eddington ratio we use the \textsc{qsosed} model to generate the corresponding spectral energy distribution assuming zero spin, an electron temperature for the hot Comptonisation component of 100\,keV, a dissipated corona luminosity at Eddington ratio of $2\%$, an electron temperature for the warm Comptonisation component of 0.2\,keV, a spectral index of the warm Comptonisation component of 2.5 and a reflection albedo for the reprocessed flux of 0.3. The observed $\log{L_\nu(\rm 2500 \mathring{A})}$ luminosity of a QSO is used to scale the model spectral energy distribution and estimate the corresponding  $L_X(\rm 2\,keV)$. For each QSO selected in a given Eddington ratio, $\log{L_\nu(\rm 2500 \mathring{A})}$ and redshift interval we can therefore estimate its model predicted X-ray luminosity. The shaded regions in Figure \ref{fig: LUV_z_bins_ledd_qsosed} correspond to the 84th percentiles  around the median of the model predicted $L_X(\rm 2\,keV)$ of the sample. We choose to show results only for the zero spin case. For a maximal spinning black hole the \textsc{qsosed} predicts a higher $L_X(\rm 2\,keV)$ at fixed black hole mass and Eddington ratio. 

The results of Figure \ref{fig: LUV_z_bins_ledd_qsosed} show a significant discrepancy between observations and model predictions. The latter show a systematic drop of the $\rm 2\,keV$ luminosity with increasing Eddington ratio as a result of the decreasing hot X-ray corona size with $\lambda_{Edd}$. The amplitude of this trend is typically not mirrored by the observational measurements, which only mildly, if at all, vary across nearly $1\,\text{dex}$ of Eddington ratio. As a result the difference between the observations and the \textsc{qsosed} model predictions increases toward lower $\lambda_{Edd}$.

%% file: Discussion_v1.tex
\section{Discussion}\label{sec:discussion}
In this work we revisit the relation between the X-ray and UV luminosities of QSOs using a sample of unprecedented size  (total of 136{,}745 unique SDSS DR16Q QSOs) with X-ray observations from either {\it XMM-Newton} or the eROSITA All Sky Survey DR1. This unique dataset is analysed by a new statistical methodology based on hierarchical Bayesian modeling, which effectively leverages the available information for improved inference. A key feature of this method is its consistent and homogeneous treatment of X-ray detections and upper limits, achieved by accurately modeling the Poisson nature of the X-ray observations. The statistical approach is also sufficiently versatile to allow both parametric and non-parametric modelling of the X-ray--to--UV correlation to measure its scatter, assess potential redshift evolution and explore the dependence on physical parameters such as the Eddington ratio. The combination of large dataset and novel methodology enables a comprehensive study of the X-ray--to--UV relation without having to apply X-ray signal-to-noise cuts that could bias the results.

First we confirm a correlation that is in qualitative agreement with literature results \citep[e.g.][]{Steffen2006, LR2016, Nanni2017, Timlin2021, Bisogni2021, Rankine2024}, i.e. an increase of the $\log L_X({\rm 2\,keV})$ toward higher $\log{L_\nu(\rm 2500 \mathring{A})}$. This relation can be described by a log-linear function (see Equation \ref{eq:model-flavour-3}) with a slope that is less that unity (see Table ~\ref{tab:parameters}). The implication is that the disc emission component of QSOs becomes increasingly more dominant relative to the X-ray corona radiative output with increasing accretion luminosity. The sublinear correlation, according to the analytical model of \citet{Arcodia2019}, may be attributed to the role of magnetic stresses and their impact on coronal-heating efficiency, which decreases with increasing accretion rate and hence, disc luminosity. Quantitatively, the inferred correlation for our full sample has a normalisation that lies at the low end of previous results (see Figure \ref{fig:comp_literature}). We believe this is related to the homogeneous treatment of both X-ray detections and upper limits afforded by the hierarchical Bayesian approach.

\subsection{Redshift Evolution of the X-ray/UV correlation}

The redshift dependence of the X-ray--to--UV correlation has been debated in the literature motivated by claims that the this relation can be used to constrain cosmological parameters \cite[e.g.][]{LR2015}. Although most prior studies argue for a non-evolving X-ray--to--UV correlation \citep[e.g.][and references therein.]{LR2015, Salvestrini2019, LR2020},  there are claims for the opposite \citep[][]{Kelly2007, Wang2022_copulas, Khadka2022, Rankine2024}. For example, \cite{Li2021} report a systematic change of the slope of the X-ray--to--UV correlation to $z=3$ for the sample of \cite{LR2016} that is of similar amplitude to that found by our analysis at the top panel of Figure \ref{fig:params_z}. They suggest however, that this may because of the narrow luminosity baseline of their redshift subsamples. This effect is reduced in the case of our QSOs that are culled from a much larger parent sample effectively allowing a reasonably wide luminosity baseline at fixed redshift (see Figure \ref{fig:Stan_lines_z}). Moreover, the simulations presented in Appendix \ref{ap:validation} show that the luminosity baseline of individual redshift bins does not bias the slope inference. \cite{Rankine2024} argue for an increasing normalisation toward higher redshift at constant slope, in contrast to our results. Direct comparison between our inferred X-ray--to--UV log-linear relations and theirs at the same redshift intervals in Figure \ref{fig: Compare_rankine} shows that our inferred evolution is significantly milder than that found by \cite{Rankine2024}. The simulations presented in Appendix~\ref{ap:validation} show that if the DR16Q QSO population followed the evolving pattern found by \cite{Rankine2024}, our methodology would recover it.  \cite{Wang2022_copulas} model the QSO sample of \citet{LR2020} with copulas and conclude that adding a redshift-dependent term to the normalization of the standard log-linear form is favoured by their analysis. However, their predicted evolution  has an amplitude of $\Delta{L_X(\rm 2\,keV)} \approx 1.6$\,dex in normalization at redshift $z=3$ (relative to $z=0$), much stronger than that found by our analysis or \cite{Rankine2024} . We emphasise that a model with varying slope of the log-linear X-ray--to--UV correlation has not been tested by \cite{Wang2022_copulas}. \cite{Khadka2021b} and \cite{Khadka2022} performed a multiparametric model fitted to separate samples including that of \citet{LR2020} to conclude that there is both redshift and cosmology dependence of the X-ray--to--UV correlation.  The amplitude and direction of the evolution is not straightforward to extract from the \citet{Khadka2022, Khadka2021b} results to compare with our analysis. 


\begin{figure*}
    \centering
    \includegraphics[width=1.0\linewidth]{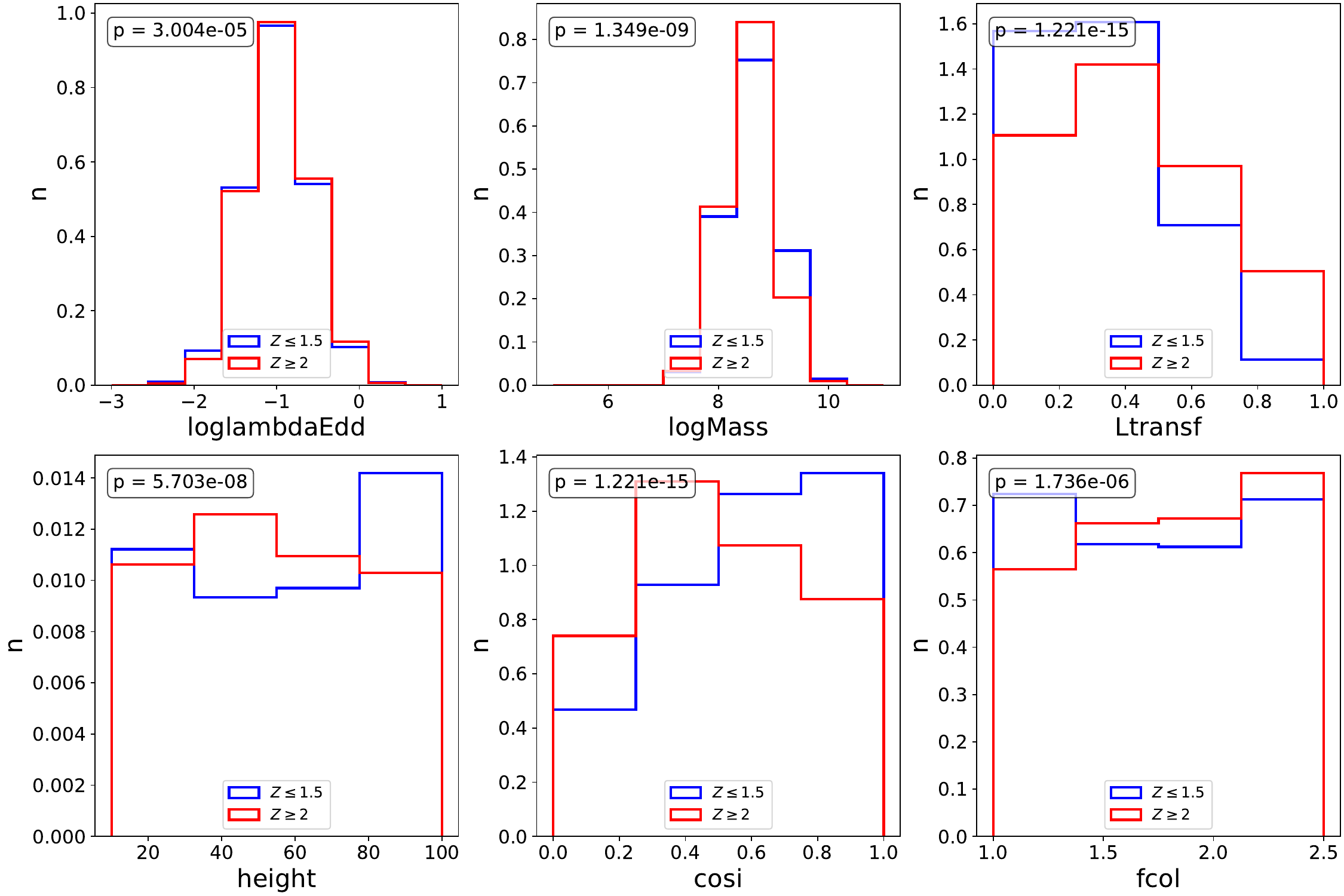}
    \caption{Histograms of the  {\sc kynsed} model \citep{Dovciak2022} parameters $\rm logMass$, $\rm loglambdaEdd$, $\rm L_{\rm transf}$, corona height, $\rm cosi$, and $\rm f_{col}$ for the SEDs from \citet{Kammoun2025}, which are consistent with our inferred $\log{\rm L_X\,(2keV)} - \log{\rm L_\nu(2500\mathring{A})}$ lines (see Fig. \ref{fig:Stan_lines_z}) for $z\leq1.5$ (blue) and $z\geq2$ (red). These are the parameters which are found to differ between the SEDs following the $\log{\rm L_X\,(2keV)} - \log{\rm L_\nu(2500\mathring{A})}$ correlation as inferred from the lower ($z\leq 1.5$)- and the higher ($z\geq 2. $)- redshift regime, at a $4\sigma$ confidence level. The p-values of the Kolmogorov-Smirnov two-sample tests for each parameter are provided within each panel. }
    \label{fig:Kammoun}
\end{figure*}

We further attempt to understand what is the possible physical interpretation of the inferred redshift evolution within the framework of the {\sc kynsed} model \citep{Dovciak2022}, a relativistic spectral model that simulates the full spectral energy distribution of accreting black holes. It combines a physically motivated accretion disc emission with a lamp-post corona geometry for the X-ray source \footnote{Although the models employed in this work assume a lamp-post X-ray corona, we note that determining the geometry of the X-ray corona remains a non-trivial task. Different studies, based on various physical mechanisms (e.g. emissivity profiles, variability, X-ray polarization), favour either compact or extended coronae with different geometries. For a detailed discussion, we refer the reader to the review of \citet{Laha2025} and references therein.}, incorporating full general relativistic ray tracing in Kerr spacetime. The model accounts for effects such as color temperature correction, disc irradiation, and relativistic light bending, allowing consistent modeling of both the optical/UV thermal emission and the X-ray corona, as well as their interplay. Free parameters of this model include the Eddington ratio, black hole mass, corona height relative to the disc plane, X-ray power-law spectral index, high energy cutoff of the X-ray spectrum, disc color correction, fraction of energy dissipated to the corona, the inclination angle of the accretion disc relative to the line-of-sight.

For our purposes we use the {\sc kynsed} SEDs  generated by \citet{Kammoun2025}  to explore the {\sc kynsed} physical parameter space that is consistent with the X-ray--to--UV correlations inferred at different redshift intervals in Figure~\ref{fig:Stan_lines_z}. We start from the total sample of "physical" SEDs of \citet{Kammoun2025}, i.e. those models with a coronal radius smaller than the height of the corona. From these "physical" SEDs we further select via importance sampling a total of 50,000 to match the observed Eddington and black hole mass distributions of the SDSS QSO sample. Finally, from these 50,000 SEDs we select those that are within $2\sigma$ of the inferred lines of Fig.~\ref{fig:Stan_lines_z}. For the selected SEDs we explore variations with redshift of the corresponding {\sc kynsed} physical parameters. For a clearer visualization and comparison, we group the SEDs that are in agreement with the correlations at $z\leq1.5$ (low redshift) and $z\geq2.0$ (high redshift). We choose non-neighboring redshift intervals to maximise the difference between the corresponding X-ray--to--UV correlations and get a clearer picture of any variations of the {\sc kynsed} model parameters. Figure~\ref{fig:Kammoun} displays the six parameters that show the most significant systematic differences between the two redshift subsamples. These were selected based on a KS two-sample test, with differences found to be significant at a confidence level exceeding $4\sigma$. These are the black hole mass, Eddington ratio, the fraction of the total disc luminosity dissipated on the corona, $\rm L_{\rm transf}$, the height of the corona above the BH in units of gravitational radii ($R_g$), the inclination angle of the accretion disc with respect to the line-of-sight, $\rm cosi$, and the color correction factor $\rm f_{col}$, accounting for the deviation of the disc emission from the black body due to illumination from the corona. 

We note that the observed variations with redshift of the black hole mass and Eddington ratio distributions are imposed by the Sloan QSO targeting selection effects. Nevertheless, these variations may play a role on the inferred X-ray--to--UV correlation. For example, the weak trend in Figure \ref{fig:Kammoun} of  smaller black hole masses toward higher redshift would tend to increase the normalisation of the X-ray--to--UV correlation for higher redshift QSOs. This is a consequence of the increasing accretion disc with decreasing black hole mass. However, in the context of the {\sc kynsed} model, any redshift dependence of the  $M_{BH}$ and $\lambda_{Edd}$ distributions cannot fully explain the observed redshift variations in the X-ray--to--UV correlation. Indeed, Figure \ref{fig:Kammoun} shows that additional {\sc kynsed} parameters are required to change systematically with redshift, i.e. $\rm f_{col}$, $\rm L_{\mathrm{transf}}$ increase, while the coronal height and the inclination angle, $\rm cosi$, decreases with redshift.


In the context of {\sc kynsed}, a higher $\rm f_{col}$ corresponds to an accretion disc with a higher effective temperature, shifting the peak of the optical/UV emission toward higher energies. This leads to a decrease in $L_\nu(\rm 2500\mathring{A})$ at fixed $L_X(\rm 2\,keV)$ and could produce the observed flattening of the X-ray--to--UV correlation at higher redshift. In the same direction acts the increasing coronal height at the lower-redshift regime, which primarily affects the flux from the disc at $2500\rm \mathring{A}$. Specifically, an increased coronal height increases the overall temperature of the disc because of the higher level of illumination by the corona. Larger heights toward lower redshift would therefore tend to steepen the X-ray--to--UV correlation. Additionally, the observed shift of the inclination parameter $\rm cosi$ toward lower values at high redshifts implies a preference for more edge-on disc orientations. This result is counter-intuitive, as the inclination angle relative to the line-of-sight is not expected to evolve with redshift. Nonetheless, this trend also decreases $L_\nu(\rm 2500\,\mathring{A})$ because of the anisotropic disc emission, thereby also contributing to a flattening of the X-ray--to--UV correlation. Finally, the fraction of dissipated energy transferred to the X-ray corona, $\rm L_{\mathrm{transf}}$, shifts toward higher values in the higher-redshift regime. The net effect is higher X-ray luminosity output at fixed $L_\nu(\rm 2500\mathring{A})$ for QSOs at higher redshift. $\rm L_{\mathrm{transf}}$ is inversely proportional with the X-ray bolometric correction (see \citealt{Kammoun2025}), thereby suggesting systematic variations of this parameter with cosmic-time. Although current studies do no report evidence for a redshift evolution of the X-ray bolometric correction \citep[e.g.][]{Duras2020}, this result highlights the need for carefully designed future experiments to test this potential dependence based on anticipated deeper multiwavelength data from new-generation and future missions (e.g., Vera Rubin Observatory, \citealt{Ivezic2019}; Euclid, \citealt{Mellier2025}; newATHENA X-ray observatory, \citealt{Cruise2025}; PRobe Far-Infrared Mission for Astrophysics -- PRIMA, \citealt{Glenn2023}). It is also worth emphasising that the predicted variations of the {\sc kynsed} model parameters with redshift in Figure \ref{fig:Kammoun}, although statistically significant, are mild with the corresponding distributions being broad.  

\subsection{Scatter of the X-ray/UV correlation}\label{sec:disc-scatter}

Another important parameter entering this analysis is the scatter of the X-ray--to--UV correlation. Its origin has been the objective of numerous works, particularly those aiming at  utilizing the X-ray--to--UV relation as a cosmological tracer \citep{LR2016,LR2017, LR2019, Bisogni2021}. These works demonstrate that a strict filtering of the sample to control for e.g. dust reddening, X-ray absorption or low X-ray signal-to-noise sources \citep[e.g.][]{LR2016, Signorini2024}, can reduce the logarithmic scatter from 0.4 to about 0.24\,dex \citep[e.g.,][]{LR2016,LR2017}. Applying such selections, although crucial for obtaining an appropriate clean cosmological sample with `typical' properties, is beyond the scope of this work\footnote{For example, following the criteria of \cite{LR2016}, we find that limit our sample to QSOs with X-ray $S/N > 5$ leads to a decrease of the inferred scatter by about $0.1$\,dex.}. Our goal is to investigate the correlation and its scatter from an inclusive point of view by incorporating the different states of accretion and the possible differences in the underlying mechanism linking the disc and the corona among QSOs . 

The "intrinsic" dispersion constrained by our log-linear parametric model is $\sigma\approx 0.39$ for the total sample, which is in agreement with results in the literature. Moreover, we find a systematic decrease of the dispersion with redshift, a trend that has also been claimed in previous studies \citep[e.g.,][]{Li2021, Rankine2024}. An important component of the inferred dispersion could be variability. The QSOs in our sample have non-simultaneous UV/optical and X-ray observations that are separated by up to about a decade at rest-frame (assuming a mean QSO redshift of about 1.5). Such time differences could introduce scatter in the inferred X-ray--to--UV correlation. We assume that the dominant factor are the X-ray flux variations, which typically show larger amplitude changes at fixed timescale compared to the UV/optical \cite[e.g.][]{Arevalo2023, Petrecca2024, Georgakakis2024}. Only about 10\% of our QSOs have multiple {\it XMM-Newton} observations taken at different epochs, which are averaged in our analysis (see section \ref{sec_Data}) thus rendering the impact of X-ray variability on the scatter of the X-ray--to--UV correlation smaller. For most Sloan QSOs only single epoch X-ray data are available, which are prone to fluctuations because of the stochastic nature of the X-ray variability. We quantify this effect using the X-ray structure function of \cite{Georgakakis2024}. We estimate a logarithmic variability scatter in the observed X-ray flux and hence the X-ray--to--UV correlation, of about $0.3$\,dex for the rest-fame time scale of 10\,yr (see above). Under the assumptions above, the residual intrinsic dispersion of the X-ray--to--UV correlation is estimated to be about $0.27$\,dex. In the variability scenario the redshift evolution of the scatter may be related to the fact that lower black hole masses at DR16Q are mainly observed at lower redshifts, and hence, higher levels of X-ray variability \citep[e.g.][]{McHardy2004, Ponti2012, Akylas2022, Georgakakis2024}.

Contamination of the AGN UV/optical light by stellar emission from the host galaxy may also contribute to the scatter of the X-ray/UV correlation. This effect likely becomes more important toward lower redshifts and lower accretion luminosities. Our analysis includes corrections to $\log{L_\nu(\rm 2500 \mathring{A})}$ that account for the stellar component of QSO hosts to $z=0.8$ \citep{Ren2024}. At higher redshift the cosmological dimming of the UV/optical galaxy profile likely increases the contrast between stellar light and unresolved nuclear emission, thereby reducing any contamination biases.

The DR16Q QSO sample is limited by flux and, therefore, it is more sensitive to low $\log{L_\nu(\rm 2500 \mathring{A})}$ sources toward lower redshift. This means that the low redshift subsamples may contain QSOs over a wider range of accretion states (e.g. Eddington ratio) compared to higher redshift bins. This diversity may also reflect on the scatter of the X-ray/UV correlation and lead to the observed redshift dependence.

\subsection{Eddington ratio dependence of the X-ray--to--UV correlation}

The DR16Q QSO sample used in this study is large enough to allow investigation of the dependence of the X-ray--to--UV correlation on the Eddington ratio. Such a dependence is predicted by accretion flow models \citep[e.g.][]{Kubota_Done2019, Mitchell2023}, in which the relative sizes of the hot X-ray corona and accretion disc are physically connected and correlated with the Eddington ratio. Observations also suggest that high Eddington ratio sources may indeed be X-ray faint for their UV luminosity, although X-ray obscuration or variability may also play an important role in shaping these trends \citep[][and references therein.]{Laurenti2022, Inayoshi2024, Wang2022, Zhang2023, Brandt2000}. Instead, our analysis in
Figures \ref{fig: LUV_z_bins_ledd} and \ref{fig: LUV_z_bins_ledd_qsosed} 
show at best a mild dependence of the X-ray/UV correlation on Eddington ratios. Moreover, a striking result in Figure \ref{fig: LUV_z_bins_ledd_qsosed} is the apparent discrepancy between observations and predictions of the \textsc{qsosed} model of \cite{Kubota_Done2019}. Contrary to our observational results the model predicts a systematic drop of the $\rm 2\,keV$ monochromatic luminosity with increasing Eddington ratio $\lambda_{Edd}$, which reflects the decreasing size of the hot X-ray corona. The discrepancy between the observations and the \textsc{qsosed} model predictions is stronger for the lower $\lambda_{Edd}$ QSOs. \cite{Mitchell2023} also explored in detail the performance of accretion flow models, including \textsc{qsosed}, against the observed X-ray/UV/optical spectral energy distributions of QSOs. They claim that the \textsc{qsosed} model roughly approximates the overall spectral energy distributions of QSOs. In detail, however, this model, and in fact all accretion flow models investigated, produce too cool UV/optical spectral energy distributions compared to the observations. Our analysis demonstrates that such discrepancies are also evident in the relative normalization of the UV and X-ray spectra of QSOs. The $\alpha_{OX}$ of active black holes appears striking similar across a reasonably broad range of Eddington ratios. 

This behaviour is a consequence of the \textsc{qsosed} model assumption that the dissipated corona luminosity is a fixed fraction of 2\% of the Eddington luminosity, which in turn sets the size of the X-ray emitting corona. This introduces an anti-correlation between X-ray luminosity and Eddington ratio (see Section 5.1 of \citealt{Kubota_Done2018}), which is what is seen in Figure \ref{fig: LUV_z_bins_ledd_qsosed}. We test this assumption in Appendix~\ref{ap:LxLedd}, to find that it is not supported by the observations. Relaxing the scaling between the hot corona dissipated energy and Eddington luminosity, or equivalently enforcing a small hot X-ray corona with a size which is independent of Eddington ratio could address the discrepancy in Figure \ref{fig: LUV_z_bins_ledd_qsosed}. For example using more generic \textsc{agnsed} model of \cite{Kubota_Done2018} that allows more freedom in the choice of parameters (e.g. the hot corona size) it is possible to improve the agreement with to observations in Figure \ref{fig: LUV_z_bins_ledd_qsosed} by fixing the hot corona X-ray size to the innermost stable circular orbit (i.e. 6 gravitational radii in the case of zero spin). 

We caution that the UV/optical and X-ray observations used to construct the data points in Figure \ref{fig: LUV_z_bins_ledd_qsosed} are not simultaneous, and, therefore, variability is not accounted for in the comparison with the model predictions. In section \ref{sec:disc-scatter}, we estimate a scatter of about 0.3\,dex because of stochastic X-ray variability, i.e. the observed X-ray fluxes/luminosities lie within 0.3\,dex of the mean at the 68\% confidence level. We therefore expect an additional scatter component of that amplitude that could be added (in quadrature) to the shaded regions of the model predictions in Figure \ref{fig: LUV_z_bins_ledd_qsosed}. However, this uncertainty is not expected to be systematic, but random, and, therefore, cannot by itself explain the discrepancy between observations and model predictions in that figure. Systematic uncertainties in the estimation of black hole masses from single epoch optical spectra also affect the inferred Eddington ratios and could smooth out any trends with this parameter. For example, \cite{Mitchell2023} suggest that black hole mass systematic uncertainties of up to 0.7\,dex may be expected when using single epoch spectra. Such a large bias may swamp any intrinsic covariances between Eddington ratio and $a_{OX}$. Nevertheless, many of the panels in Figure \ref{fig: LUV_z_bins_ledd_qsosed} span 1.5\,dex in Eddington ratio and therefore one would expect any intrinsic correlations of the amplitude predicted by the \textsc{qsosed} model to leave their signatures in the inferred $L_X(\rm 2\,keV)$  of widely separated $\lambda_{Edd}$ bins. Overall, our observational results do no support a strong dependence of the $\log{L_\nu(\rm 2500 \mathring{A})}-\log{L_X(\rm 2\,keV)}$ correlation on the Eddington ratio. 

\subsection{X-ray faint population}

\begin{figure}
    \centering
    \includegraphics[width=1.0\linewidth]{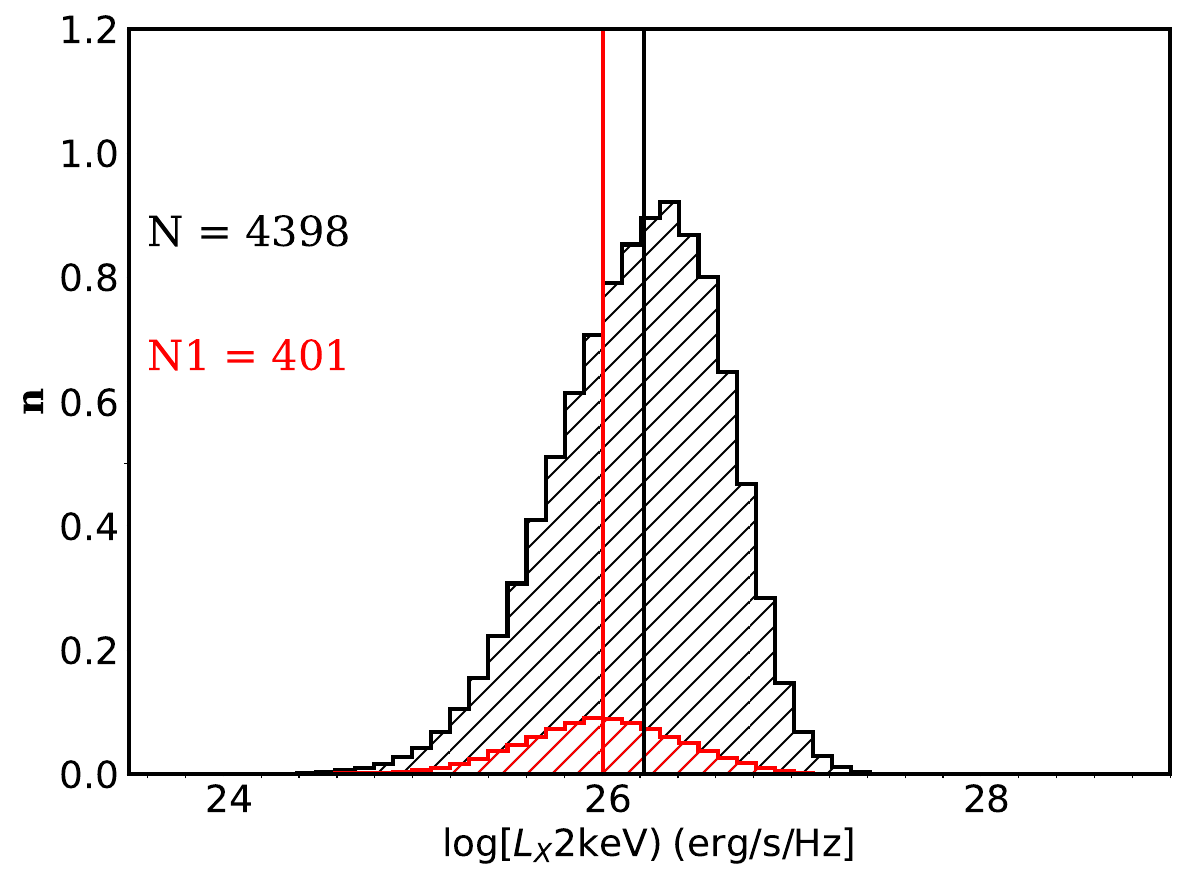}
    \caption{The posterior distribution of $\log{L_X(2\,keV)}$ for the bin $\log{L_\nu(\rm 2500 \mathring{A})} = [30.0-30.5]$ and $z = [0.5-1.0]$. The red histograms correspond to the QSOs with $\Delta(g-i) <0.3$. The vertical lines of respective colors indicate the median values of the distributions. The number of QSOs for each histogram are written in the respective color. }
    \label{fig:Dgi-example}
\end{figure}

Apart from their potential role in our results, the X-ray faint QSO population is interesting by itself for further investigation. We report a fraction of $\approx 3\%$ of QSOs with an $\Delta\alpha_{OX} < -0.3$ with respect
to the total best fit relation for the full sample and a fraction of $\approx 0.7\%$ of objects with $\Delta\alpha_{OX} < -0.38$  (X-rays underluminous for the UV luminosity by at least 1\,dex).  These fractions are lower than those reported by \citet{Pu2020}, $\sim 5.8\%$ and $2.7\%$ for objects with $\Delta\alpha_{OX} < -0.3$ and $\Delta\alpha_{OX} < -0.38$, respectively. \citet{Gibson2008} estimate that $\approx 2\%$ of their SDSS QSOs have $\Delta\alpha_{OX} < -0.38$ (X-ray faint by a factor of $\gtrsim 10$). Results based on high-redshift quasar samples indicate a substantially larger fraction of X-ray–weak QSOs compared to lower-redshift studies. In particular, \citet{Nardini2019}, analysing a sample of 30 quasars at redshift 
$z=3-3.3$, report an X-ray–weak fraction of approximately 25\%. Similarly, \citet{Zappacosta2020} and \citet{DegliAgosti2025}, using the WISE/SDSS-selected hyper-luminous (WISSH) quasar sample spanning $z\approx 2-4$, find fractions of around 30\%. It is worth noting however, that these high-redshift samples are relatively small and thus subject to low-number statistical uncertainties. The differences between our findings and the aforementioned works, can be partially justified considering the differences between their samples and the one used in this work, in terms of both size and range of physical parameters (e.g. luminosity). Constraining our sample to the redshift interval of the studies above has negligible impact on the estimated fractions. Also, the sample of \citet{Pu2020} (their parent sample A) contains about $1/3$ of non X-ray detected QSOs, while that of \cite{Gibson2008} includes X-ray detections only. Our sample is instead largely dominated by upper limits. Selecting only those QSOs that correspond to  $4.5\sigma$ X-ray detections only, results to fractions of $\sim 8\%$ and $2.8\%$ for $\Delta\alpha_{OX} < -0.3$ and  $\Delta\alpha_{OX} < -0.38$, respectively.

Attempting to understand the nature of the X-ray weak QSOs  in our sample we investigate several of their statistical parameters with respect to the general QSO population. Interestingly, they do not show any statistical differences in terms of $\lambda_{Edd}$ or $M_{BH}$. Motivated by the X-ray-weak-QSO classification of \citet{Pu2020}, we investigate a possible link between the QSO reddening, i.e. optical absorption, and X-ray faintness, which would reveal a co-presence of dust and gas. Following \citet{Pu2020} we adopt the redshift-independent reddening quantification criterion of \citet{Richards2003} and select QSOs with $\Delta(g-i)>0.3$. We find no evidence of a greater incidence of redder objects within our X-ray population than in the general population, which is about $10\%$. However, for some of the $L_\nu(2500\mathring{A}) - z$ intervals of Fig. \ref{fig:LUV_z_bins},  the inferred $\log{L_X(\rm 2\,keV)}$ is shifted towards lower values for the 'reddened' QSOs. An example is shown in Fig. \ref{fig:Dgi-example}, which plots the $\log{L_X(\rm 2\,keV)}$ distribution separately for QSOs with $(\Delta(g-i)>0.3)$. In this case,  the reddened QSOs appear to have a lower median $L_X(\rm 2\,keV)$ than their parent subsample within the specific UV and redshift bin, by a difference of $\Delta{\log{L_X(\rm 2\,keV)}} \approx 0.6 \; \rm dex$. Assuming that this luminosity difference is the effect of X-ray absorption, it would correspond to an $N_{H} \approx 1.5 \cdot 10^{21} \rm  cm^{-2}$ which corresponds to a color excess of $E(B-V) \approx 0.26$, assuming the empirical relation of \citet{Bohlin1978}. This X-ray faint population is particularly interesting and deserves further investigation.

%% file: Conclusions.tex
\section{Conclusions}\label{conclusions}

This work revisits the X-ray--to--UV correlation of SDSS QSOs, using as tracers of the X-ray and the UV luminsoities the $2\,{\rm keV}$ and $2500 \rm \mathring{A}$ luminosities, respectively. A robust hierarchical Bayesian methodology is employed and the correlation is examined using different parametrizations. The SDSS sample is combined with {\it XMM-Newton} archival data and eROSITA observations carried out in the first six months of the SRG/eROSITA all-sky survey (eRASS1).


Our findings confirm that the X-ray--to--UV correlation can be described by a power-law with index less than unity, indicating that the SED of quasars becomes increasingly disc-dominated with higher optical/UV luminosity. While this is qualitatively consistent with previous studies, our results lie at the lower end of the X-ray--to--UV ratio range reported in the literature. 

We explore the redshift evolution of the correlation using different parameterizations and find robust evidence for a mild ($\approx 0.3-0.4 \text{\,dex}$) but systematic trend. The evolution is luminosity-dependent: while the most luminous quasars show minimal evolution, those with lower optical/UV luminosities display systematically higher X-ray luminosities with increasing redshift. Moreover, the intrinsic scatter of the correlation decreases with redshift, in line with previous results reported in the literature.

The observed evolution of the correlation, if physical, suggests a change in accretion processes and in the interaction between the fundamental components of the accretion flow over cosmic time. We interpret this result within the framework of the {\sc kynsed} model of \citet{Dovciak2022, Kammoun2025}. In this context, the direction of the evolution is consistent with a disc increasingly dominated by scattering and a higher fraction of the dissipated energy being transferred to the X-ray corona, which may suggest a redshift evolution of the X-ray bolometric correction.

The inferred evolution is systematic and remains robust across all complementary models used in this study, while the observations favour the model of our analysis where both the slope and the normalization of the correlation are redshift dependent. Deeper X-ray and UV observations are needed to confirm these trends by extending the analysis to fainter QSOs. Such datasets include the DESI-DR1 \citep{DESI2025} and the stacked eRASS1-5 observations.

Finally, the size of the sample enables an investigation of potential dependencies of the correlation on the physical properties of quasars. We find no evidence for a dependence on black hole mass. We conduct a detailed analysis of the correlation’s dependence on accretion power, parametrized by the Eddington ratio, as predicted by several recent accretion models. However, our results indicate that any such dependence is weak. Inversely, robust physical models coupled with large samples such as the one presented here could be used to assess the systematic uncertainties in the black hole physical parameters, such as single-epoch virial BH mass estimates.

%% file: Appendix_A.tex
\begin{appendix}

\section{Validation of the SED fitting}\label{ap:SED_validation}

In this section we compare the $L_{2500\mathring{A}}$ derived via the CIGALE using the setup shown in Table \ref{t1} with an alternative model-independent approach for measuring the same quantity. 

\begin{figure}
   \centering
   \includegraphics[scale=0.6]{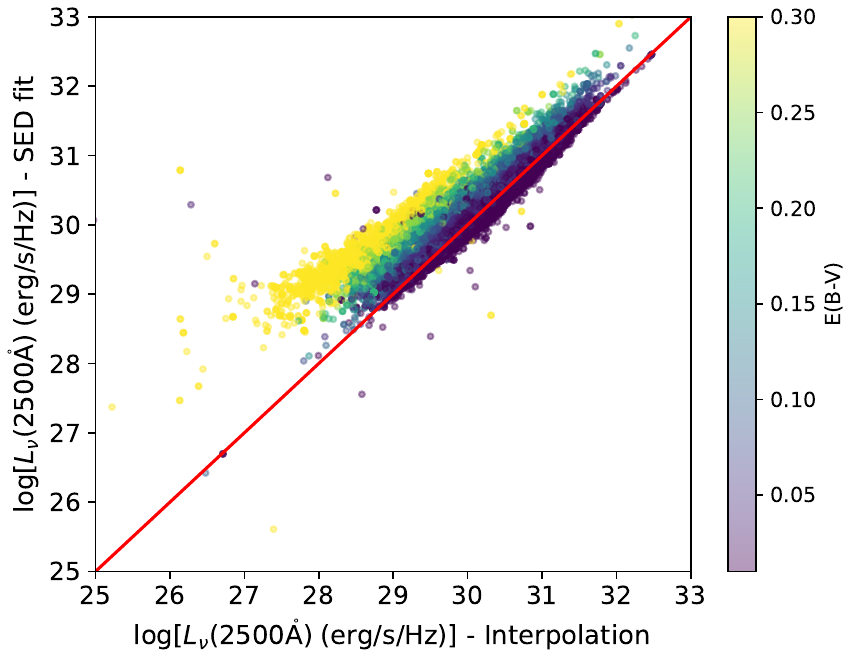}
   \includegraphics[scale=0.6]{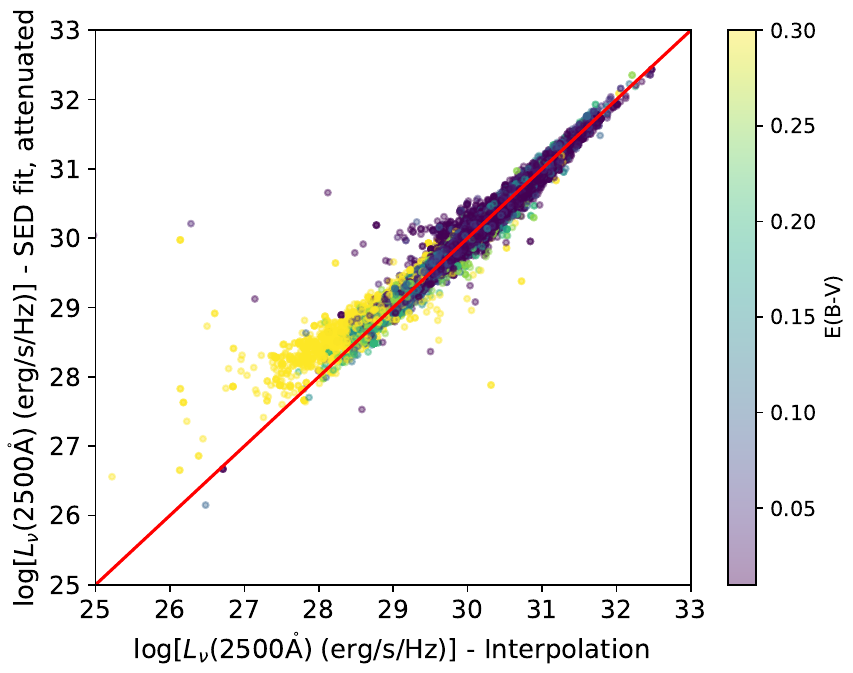}
   \caption{Comparison of $L_\nu(\rm 2500\mathring{A})$ as calculated by the interpolation procedure (horizontal axis) and CIGALE (vertical axis). The top panel plots the the intrinsic $L_\nu(\rm 2500\mathring{A})$ returned by CIGALE. The bottom panel shows the CIGALE extincted luminosity density (i.e. not corrected for dust). The level of extinction of each datapoint is expressed by the CIGALE polar direction $E(B-V)$ parameter and is colour-coded as shown in the colourbar. The red solid line is the one-to-one relation.}
    \label{Comparison_SED_INTERP}%
    \end{figure}

This method estimates $L_\nu(\rm 2500\mathring{A})$ by linearly interpolating between the observed flux densities in photometric bands that bracket the redshifted wavelength $2500{\rm \mathring{A}}\cdot(1+z)$, where $z$ is the redshift of the QSO under consideration. The interpolated flux density can then be converted to rest frame and corrected for the luminosity distance of the QSO to determine the corresponding $L_\nu(\rm 2500\mathring{A})$. The fluxes used in the linear interpolation are corrected for Galactic extinction and for any extinction, $E(B-V)$, intrinsic to the QSO, e.g. from the host galaxy. Figure \ref{Comparison_SED_INTERP} (upper panel) compares the interpolation-based 2500\AA\, luminosity densities (horizontal axis) with the intrinsic ones (i.e. extinction corrected) estimated by CIGALE (vertical axis). The bulk of the population scatter around the one-to-one line but there are sources that deviate toward higher CIGALE estimated intrinsic luminosity densities. These are the sources for which CIGALE finds evidence for non-zero extinction and therefore returns an intrinsic, dust-corrected, $L_\nu(\rm 2500\mathring{A})$. The latter point is demonstrated at the bottom panel of  Fig. \ref{Comparison_SED_INTERP}, where extinction is applied back to the CIGALE results to make them directly comparable to those from the interpolation approach. The overall scatter decreases and the populations lies on the one--to--one relation.  
 
We select the SED-fitting as a more uniform approach for the full dataset because it is based on physical models and is able to provide an estimate for the $E(B-V)$ parameter, allowing us to correct for the host galaxy extinction. 


%

\begin{table*}
\caption{\label{t1} Skirtor2016 and Redshifting modules with parameter grid used for SED fitting.}
\centering
\begin{tabular}{lllp{0.4\textwidth}}
\hline\hline
\textbf{Module} & \textbf{Parameter} & \textbf{Values} & \textbf{Description} \\
\hline
\multicolumn{4}{l}{\textbf{Skirtor2016 (Input Parameters)}} \\
\texttt{skirtor2016} & \texttt{agn.t}           & $[3.0, 7.0, 10.0]$  & Average edge-on torus optical depth at 9.7 microns. \\
                     & \texttt{agn.pl}          & $[1.0]$   & Power-law exponent that sets the radial gradient of dust density. \\
                     & \texttt{agn.q}           & $[1.0]$   & Index that sets dust density gradient with polar angle. \\
                     & \texttt{agn.oa}          & $[40]$ (deg) & Angle between the equatorial plane and the edge of the torus. \\
                     & \texttt{agn.R}           & $[20]$ & Ratio of outer to inner torus radius (\(R_\text{out}/R_\text{in}\)). \\
                     & \texttt{agn.Mcl}         & $[0.97]$ & Fraction of total dust mass inside clumps. $0.97$ means $97\%$ of total 
                     mass is inside the clumps and $3\%$ in the interclump dust. \\
                     & \texttt{agn.i}           & $[10, 30]$ (deg)  & Viewing angle: \(i=[0,90^\circ - \text{oa}]\) (face-on);  \(i=[90^\circ - \text{oa},90^\circ]\) (edge-on). \\
                     & \texttt{agn.disktype}    & $[1]$ & Disk spectrum: 0 for the regular Skirtor spectrum, $1$ for the Schartmann (2005) spectrum. \\
                     & \texttt{agn.delta}     & $[-0.36]$   & Power-law of index $\delta$ modifying the optical slop of the disk. Negative values make the slope steeper where as positive values make it shallower. \\
                     & \texttt{agn.fracAGN}     & $[0.999]$   & Fraction of AGN IR luminosity to total IR luminosity. \\
                     & \texttt{agn.lambdafracAGN}     & $[0.999]$   & Wavelength range in microns where to compute the AGN fraction. \\
                     & \texttt{agn.law}         & $[0]$       & Extinction law of polar dust: 0 (SMC), 1 \citep{Calzetti2000}, or 2 \citep{Gaskell2004}. \\
                     & \texttt{agn.EBV}         & $[0.01, 0.02, 0.05, $   & E(B-V) for extinction in the polar direction. \\
                     &         & $..., 0.25, 0.3, 0.4,...,1.]$   & \\
                     & \texttt{agn.temperature} & $[100]$ K & Temperature of the polar dust. \\
                     & \texttt{agn.emissivity}  & $[1.6]$   & Emissivity index of the polar dust. \\
\hline
\multicolumn{4}{l}{\textbf{Skirtor2016 (Output Parameters)}} \\
\texttt{skirtor2016} & \texttt{agn.disk.luminosity}    & Derived value & The observed AGN disc luminosity (might be extincted). \\
                     & \texttt{agn.dust.luminosity}    & Derived value & The observed AGN dust re-emitted luminosity. \\
                     & \texttt{agn.luminosity}         & Derived value & The sum of \texttt{agn.disk.luminosity} and \texttt{agn.dust.luminosity}. \\
                     & \texttt{agn.intrin.Lnu.2500A}   & Derived value & The intrinsic AGN \(L_\nu\) at 2500 \(\text{\AA}\) at viewing angle = \(30^\circ\). \\
                     & \texttt{agn.accretion.power}    & Derived value & The intrinsic AGN disk luminosity averaged over all directions. \\
\hline
\multicolumn{4}{l}{\textbf{Redshifting (Parameters)}} \\
\texttt{redshifting} & \texttt{redshift}       & ... & Input redshift of the source. \\
                     
\hline
\end{tabular}
\end{table*}

\section{Validation of Bayesian Inference Methodology}\label{ap:validation}

In this section we test and validate the hierarchical Bayesian methodology described in Section \ref{sec_Method} for inferring the relation between $L_X(\rm 2\,keV)$ and $L_\nu(\rm 2500\mathring{A})$ of QSOs. 

The starting point of this exercise is the DR16Q QSO sample, i.e. a catalogue of redshifts and 2500\,\AA\; monochromatic luminosities estimated in Section \ref{sec_Data}. For each QSO in the sample we simulate realistic X-ray photometric observations (both XMM and eROSITA) by adopting a parametric $L_X({\rm 2\,keV}) - L_\nu(\rm 2500\mathring{A})$ correlation.  The methodology of Section \ref{sec_Method} is then applied to these mock observations to infer the relation between 2\,keV and 2500\,\AA\;  monochromatic luminosities. For our testing and validation purposes, we use two different input relations from the literature. The first is the one derived by \cite{Lusso2010} that is independent of redshift, reading

\begin{equation}\label{eq:L2-L2500-lusso}
\log L_X({\rm 2\,keV}) = 0.76 \cdot \log L_\nu({\rm 2500\mathring{A}}) + 3.508.
\end{equation}

\noindent The second scaling relation that we use depends on redshift and allows us to explore the sensitivity of our analysis to evolutionary effects. We start from the redshift dependent $\alpha_{ox}$ vs $\log L_\nu({\rm 2500\mathring{A}})$ correlation reported by \citet[][see their Table 3]{Rankine2024}, which after some algebra can be transformed to

\begin{equation}\label{eq:L2-L2500-rankine}
\log L_X({\rm 2\,keV}) = 0.31 \cdot \log L_\nu({\rm 2500\mathring{A}}) + 0.41\cdot z + 15.88,
\end{equation}
    
\noindent where $z$ is the redshift. Equations \ref{eq:L2-L2500-lusso}, \ref{eq:L2-L2500-rankine} are used to create two independent mock QSO samples with and without redshift evolution, respectively. For both samples, we assume a scatter of 0.4\,dex around the mean $\log L_X({\rm 2\,keV}) - \log L_\nu(\rm 2500\mathring{A})$ correlation of Equations \ref{eq:L2-L2500-lusso}, \ref{eq:L2-L2500-rankine}. 

For a given QSO with available X-ray photometry from either XMM or eROSITA, redshift $z$ and UV monochromatic luminosity $L_\nu({\rm 2500\mathring{A}})$ (see Section \ref{sec_Data}) we estimate the corresponding $L_X(\rm 2\,keV)$ via Equations \ref{eq:L2-L2500-lusso} or \ref{eq:L2-L2500-rankine}. This is then converted to flux in either the 0.2-2\,keV band (case of XMM observations) or the 0.2-2.3\,keV spectral range (case of eROSITA observations) assuming a power-law spectral index of $\Gamma=1.9$ absorbed by the Galactic hydrogen column density in the direction of the QSO under consideration. The catalogued exposure time, encircled energy fraction, energy conversion factor and background values (either XMM or eROSITA) for the the QSO in question are then used to transform the observed flux to the Poisson photon count expectation value via Equation \ref{eq:poisson-expectation}. The latter is used to draw a Poisson deviate that represents the observed integer photon counts within the aperture. The end product of this process are simulated aperture photometry measurements which can be passed to the Bayesian methodology of Section \ref{sec_Method}. We split the simulated DR16Q sample into the same redshift intervals used for the real observations (see Section \ref{sec_Method}) with boundaries $z=[0.05, 0.5, 1.0, 1.5, 2.0, 2.5, 3.0]$. A linear relation of the form 

\begin{equation}\label{eq:sims-linear-fit}
\log L_X(\rm 2\,keV) = A \cdot\log L_\nu(\rm 2500\mathring{A}) + b,
\end{equation}

\noindent with Gaussian scatter $\sigma$ around the mean is fit to the mock data set within each of the redshift intervals above. 

\begin{figure}
   \centering
   \includegraphics[width=1.0\columnwidth]{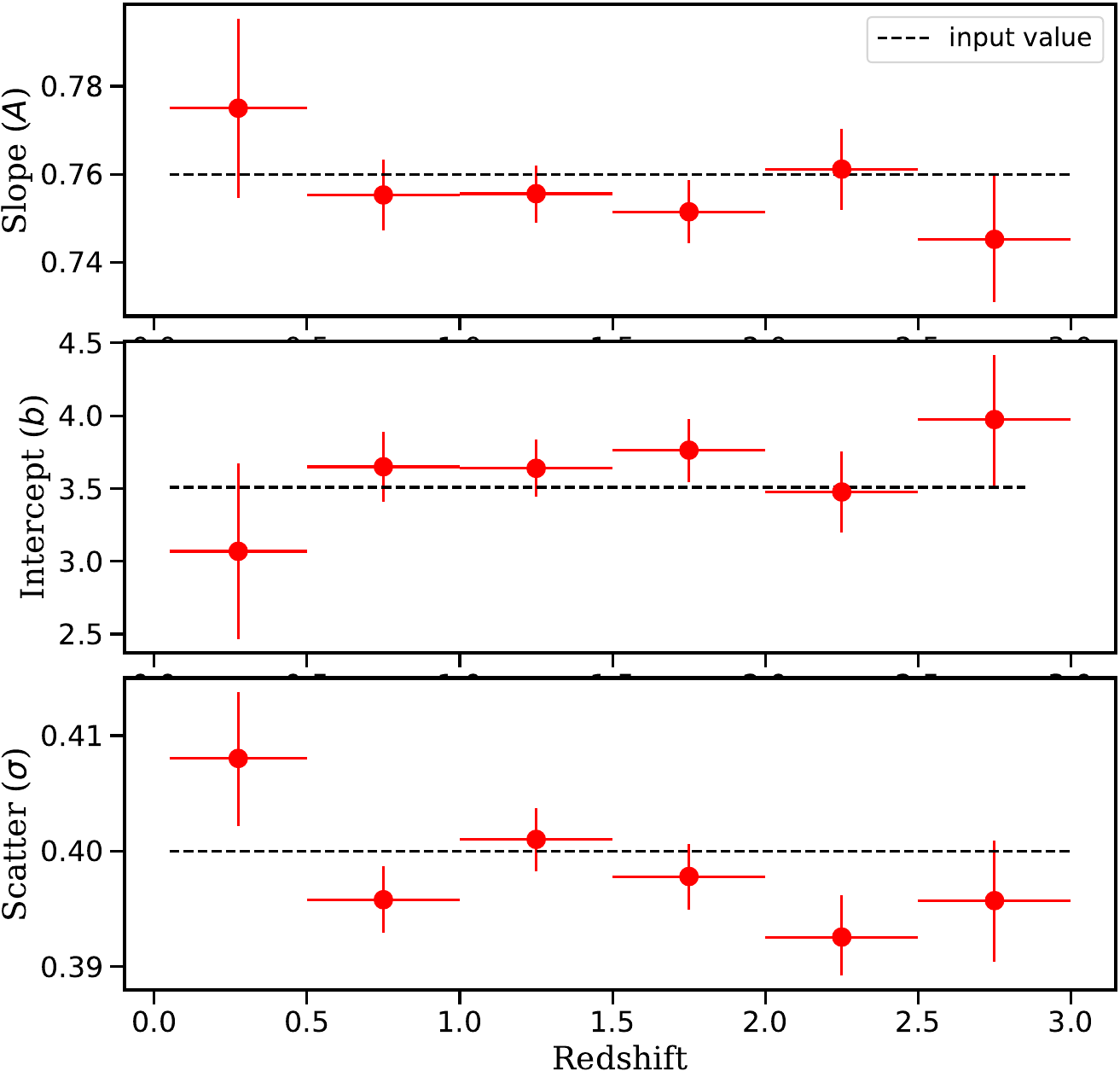}
   \caption{Inferred parameters of the linear  relation between $\log L_X(\rm2\,keV)$ and $\log L_\nu(\rm 2500\mathring{A})$ for simulated QSO samples selected in different redshift intervals. The simulations assume the redshift independent scaling relation of Equation \ref{eq:L2-L2500-lusso} to link the monochromatic luminosities at 2\,keV and 2500\AA. The different panels from top to bottom show the slope ($A$) and intercept ($b$) of the linear fit as well as the Gaussian scatter ($\sigma$) around the mean. In all panels the red data points are the medians of the parameter posterior distributions returned by the Bayesian methodology. The vertical errorbars correspond to the 68\% confidence interval around the median. The horizontal errorbars show the extent of the redshift interval within which mock QSOs are selected. The black dashed line in all panels shows the input value of the parameter used to construct the simulated X-ray photometry (see Equation \ref{eq:L2-L2500-lusso} and text for details).}\label{fig:appendix_sim_noz}
    \end{figure}

In the case of the non-evolving $\log L_X({\rm 2\,keV}) - \log L_\nu(\rm 2500\mathring{A})$ correlation of Equation \ref{eq:L2-L2500-lusso}, Figure \ref{fig:appendix_sim_noz} shows the inferred linear-fit parameters for the different redshift intervals of the mock QSO samples. It demonstrates that the Bayesian methodology of Section \ref{sec_Method} recovers reasonably well the input values. 

\begin{figure}
   \centering
   \includegraphics[width=1.0\columnwidth]{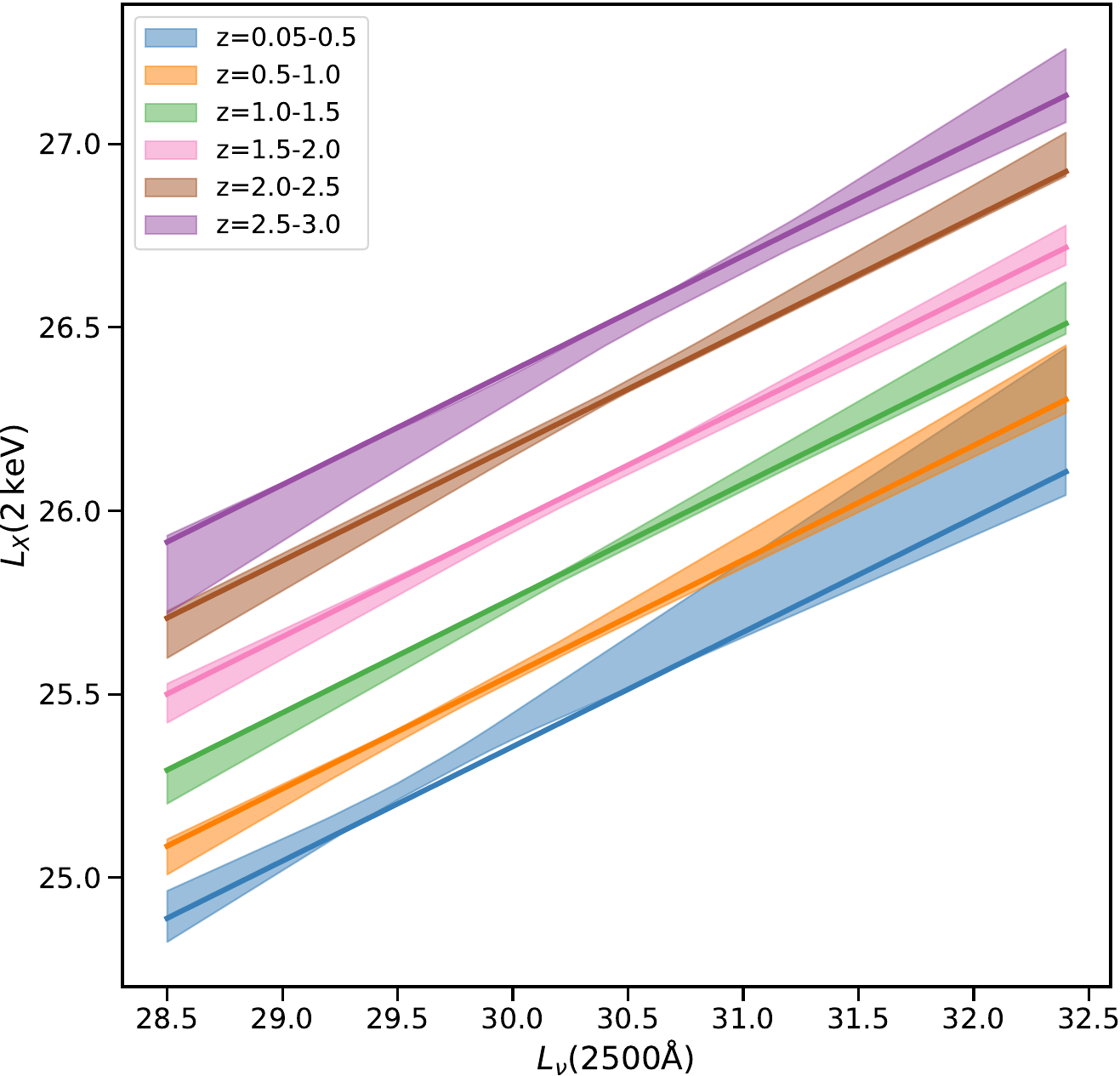}
   \caption{Inferred linear  relation between $\log L_X(\rm2\,keV)$ and $\log L_\nu(\rm 2500\mathring{A})$ for simulated QSO samples selected in different redshift intervals. The simulations assume the redshift dependent scaling relation of Equation \ref{eq:L2-L2500-rankine} to link the monochromatic luminosities at 2\,keV and 2500\AA. The shaded regions with different colours show the 68\% confidence intervals of the linear model of Equation \ref{eq:sims-linear-fit} for mock QSO sample selected at different redshift intervals as indicated in the legend. The solid thick lines of the same colour show the input $\log L_X(\rm2\,keV)$ and $\log L_\nu(\rm 2500\mathring{A})$ of Equation \ref{eq:L2-L2500-rankine} evaluated at the mean redshift of each interval.}\label{fig:appendix_sim_zevol}
    \end{figure}

Next we explore the fit to the mock QSO sample constructed by assuming redshift evolution of the  $\log L_X({\rm 2\,keV}) - \log L_\nu(\rm 2500\mathring{A})$ correlation as in Equation \ref{eq:L2-L2500-rankine}. It is emphasized that the adopted Bayesian inference model does not parametrise the redshift evolution of the sample. Instead, each redshift subsample is analysed independently to derive posteriors for the parameters (slope, intercept, scatter) of the linear relation of Equation \ref{eq:sims-linear-fit}. The results are shown in Figure \ref{fig:appendix_sim_zevol}, which plots the inferred linear model at different redshift intervals. Also shown are the input $\log L_X({\rm 2\,keV}) - \log L_\nu(\rm 2500\mathring{A})$ relations (Equation \ref{eq:L2-L2500-rankine}) evaluated at the mean redshift of each interval. The Bayesian methodology recovers the redshift evolution of the $\log L_X({\rm 2\,keV}) - \log L_\nu(\rm 2500\mathring{A})$ correlation. This is manifested by the systematic increase of the overall normalisation of the inferred linear model toward higher redshift. These results are also consistent with the evolving  $\log L_X({\rm 2\,keV}) - \log L_\nu(\rm 2500\mathring{A})$ parametrisation of Equation \ref{eq:sims-linear-fit}. 

\begin{figure}
   \centering
   \includegraphics[width=1.0\columnwidth]{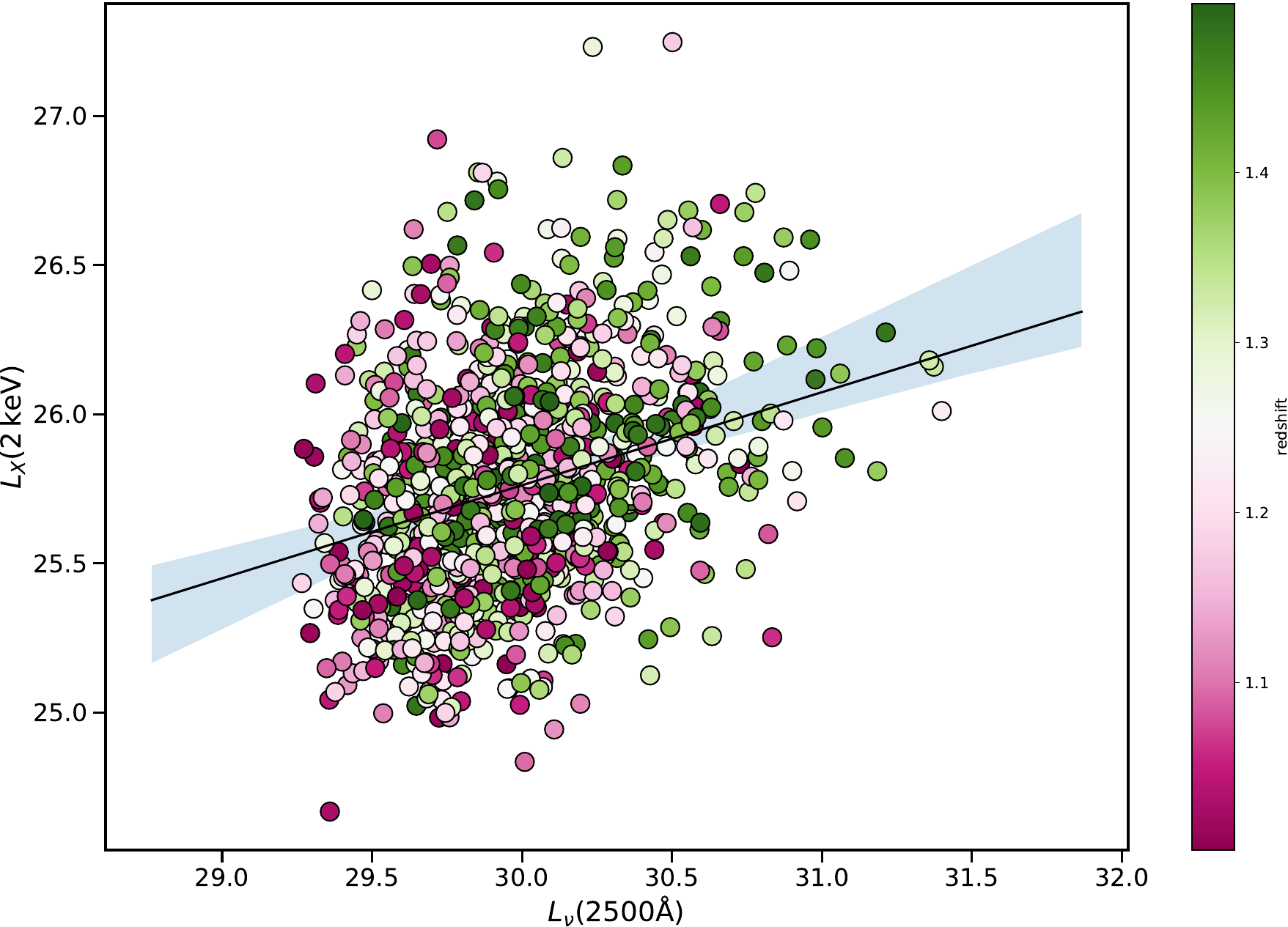}
   \caption{Demonstration of the inferred linear  relation between $\log L_X(\rm2\,keV)$ and $\log L_\nu(\rm 2500\mathring{A})$ for simulated QSOs in the redshift interval 1.0--1.5. The simulations assume the redshift dependent scaling relation of Equation \ref{eq:L2-L2500-rankine} to link the monochromatic luminosities at 2\,keV and 2500\AA. The data points represent individual QSOs in the sample colour coded by their redshift as indicated by the colourbar on the right. There is a selection effect whereby more UV luminous QSOs lie systematically at higher redshift (greener colours) whereas low luminosities are dominated by lower redshift sources (more purple colours). The shaded light-blue region shows the 68\% confidence intervals of the linear model of Equation \ref{eq:sims-linear-fit}. The solid black line shows the input $\log L_X(\rm2\,keV)$ and $\log L_\nu(\rm 2500\mathring{A})$ of Equation \ref{eq:L2-L2500-rankine} evaluated at the mean redshift of the interval.}\label{fig:appendix_sim_tilt}
    \end{figure}

There is also a systematic trend whereby the inferred slope appears steeper than the input one for all redshift intervals in Figure \ref{fig:appendix_sim_zevol}. This steepening is an observational selection effect related to the evolving X-ray/UV correlation. Toward the high redshift edge of each interval the available cosmological volume is larger. This means that within a given redshift interval the luminous QSOs will tend to lie closer to the high redshift boundary of the bin. By construction these higher redshift QSOs will also have a higher X-ray luminosity relative to those at the mean redshift of the bin (i.e. Equation \ref{eq:L2-L2500-rankine}). In contrast, the low edge of each redshift interval tends to be dominated by lower UV luminosity QSOs as a result of the flux limit of the Sloan sample. These QSOs will also be underluminous at X-rays relative to sources of similar UV luminosity that lie close to the mean redshift of the interval. These combined effects are demonstrated in Figure \ref{fig:appendix_sim_tilt} and act to introduce a tilt in the inferred linear correlation thereby leading to a mild steepening of the recovered slope in the case of a model fit given by Equation \ref{eq:sims-linear-fit}.

\section{Redshift dependent X-ray/UV parametric fit}\label{ap:Stan_zevol}

\begin{figure}
   \centering
   \includegraphics[scale=0.36]{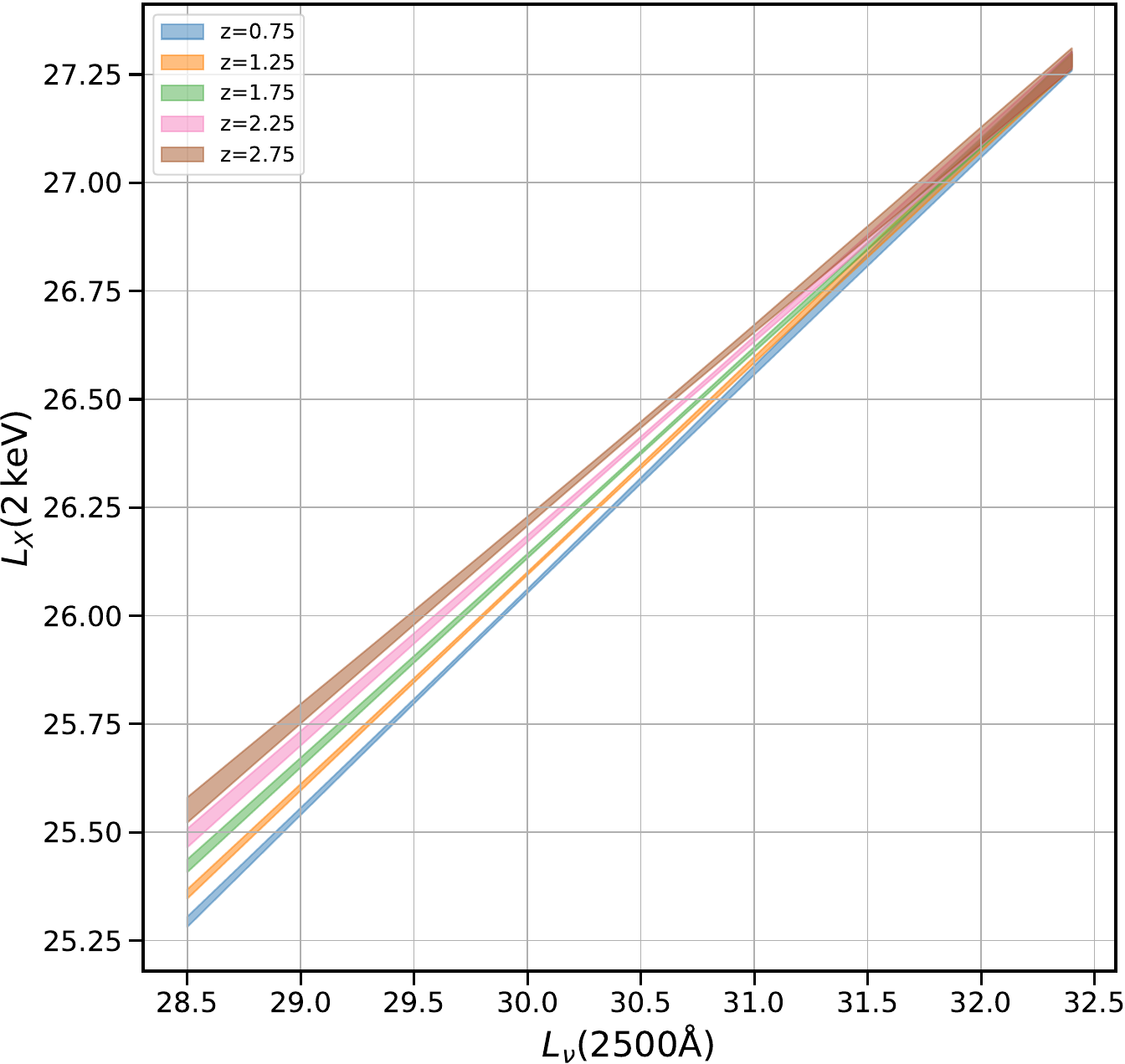}
   \caption{ Inferred $L_X({\rm 2\,keV}) - L_\nu(\rm 2500 \mathring{A})$ correlation assuming the log-linear model of Equation \protect\ref{eq:model-flavour-3} with a redshift dependent slope and intercept described by Equations \ref{eq:model-flavour-4-slope}, \ref{eq:model-flavour-4-intercept}. The light shaded regions correspond to the 90th confidence interval around the median.} \label{fig:Stan_zevol}%
    \end{figure}

Figure \ref{fig:Stan_zevol} plots the inferred X-ray/UV correlation for the parameteric log-linear model that includes a redshift dependent slope and intercept given by Equations \ref{eq:model-flavour-4-slope}, \ref{eq:model-flavour-4-intercept}, respectively.

\section{Testing the QSOSED model assumption on the corona dissipated energy}\label{ap:LxLedd}
\begin{figure}
    \centering
    \includegraphics[width=0.9\linewidth]{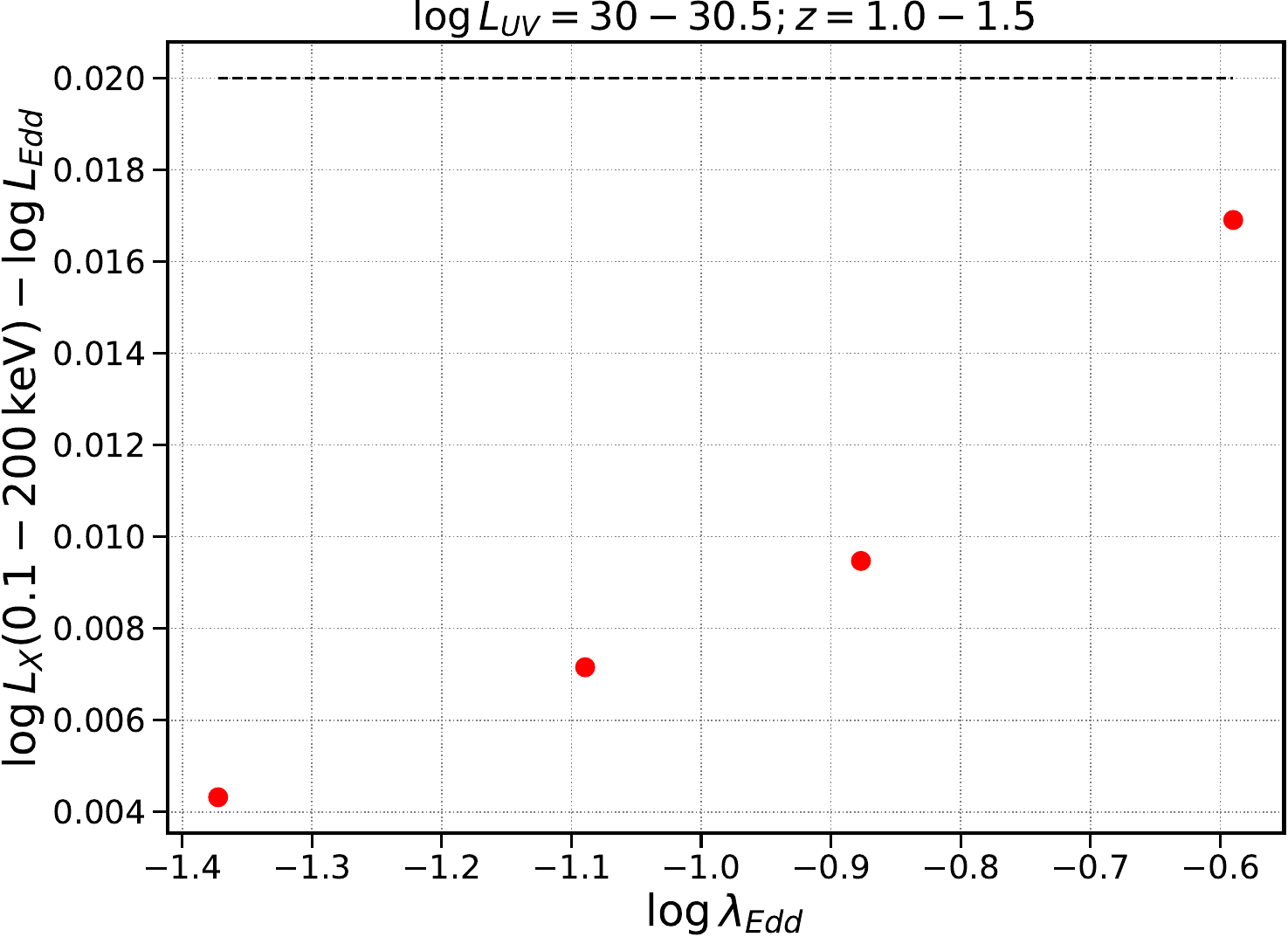}
    \caption{Ratio $L_X({\rm 0.1-200\,keV})/L_{Edd}$  as a function of Eddington ratio for QSOs with $\log [L_\nu(\rm 2500\mathring{A})/(erg\,s^{-1})]=30-30.5$ and $z=1.0-1.5$. The red points correspond to the median of the $L_X({\rm 0.1-200\,keV})/L_{Edd}$ distribution for each Eddington ratio bin. The dashed line at 0.02 represents the \textsc{qsosed} model assumption that the dissipated corona luminosity is a fixed fraction of 2\% of the Eddington luminosity.}
    \label{fig:LXLedd_lambda}
\end{figure}

In Figure \ref{fig:LXLedd_lambda} we demonstrate that the observationally inferred mean ratio $L_X({\rm 0.1-200\,keV})/L_{Edd}$ of DRQ16 QSOs decreases below 2\%  with decreasing $\lambda_{Edd}$. The $L_X({\rm 0.1-200\,keV})$ represents the total hot corona luminosity and is meant to approximated the corona dissipated energy of the \textsc{qsosed} model of \cite{Kubota_Done2019}.  It is estimated by scaling the observationally inferred $L_\nu({\rm 2\,keV})$ of each QSO to the energy interval 0.1-200\,keV assuming an X-ray photon spectral index of 1.9. Although Figure \ref{fig:LXLedd_lambda} shows results only for the subsample  with $\log [L_\nu(\rm 2500\mathring{A})/(erg\,s^{-1})]=30-30.5$ and $z=1.0-1.5$, similar conclusions are obtained for all the UV luminosity and redshift panels of \ref{fig: LUV_z_bins_ledd_qsosed}.

\end{appendix}